\def\ARXIVVERSION{}
\documentclass[preprint,12pt]{elsarticle}
\usepackage{amsmath}
\usepackage{xspace}
\usepackage{xurl}
\usepackage{booktabs}
\usepackage[dvipsnames]{xcolor}
\usepackage{listings}
\usepackage{comment}
\usepackage{caption}
\usepackage{multicol}

\usepackage{graphicx}
\usepackage{subcaption}
\usepackage{float}
\usepackage{tikz}
\usetikzlibrary{calc}

\usepackage{amssymb}
\newcounter{bla}

\journal{Computer Physics Communications}
\date{}

\newcommand{\nektarpp}{Nektar\texttt{++}\xspace}
\newcommand{\dt}{\Delta t}
\newcommand{\dtcfl}{\dt_{\mathrm{CFL}}}
\newcommand{\ndt}{N_{\dt}}
\newcommand{\tdt}{T_{\dt}}
\newcommand{\tsol}{T_{\text{sol}}}
\newcommand{\dtau}{\Delta \tau}
\newcommand{\ndtau}{N_{\dtau}}
\newcommand{\ctu}{~\text{CTU}}
\newcommand{\eqordernote}{\rev{Linear-implicit cases at $\dt \geq 20\,\dtcfl$ use equal-order spaces ($P_p = P_v = 4$), see section~\ref{sec.stability}.}}

\ifdefined\ARXIVVERSION
  \newcommand{\rev}[1]{#1}
  \newenvironment{revblock}{\par}{\par}
\else
  \newcommand{\rev}[1]{\textcolor{blue}{#1}}
  \newenvironment{revblock}{\par\color{blue}}{\par\color{black}}
\fi

\begin{document}

\begin{frontmatter}

\title{Implicit Velocity Correction Schemes for Scale-Resolving Simulations of Incompressible Flow: Stability, Accuracy, and Performance}

\author[a]{Henrik Wüstenberg\corref{author}}
\author[a,b]{Alexandra Liosi}
\author[a]{Spencer J. Sherwin}
\author[a]{Joaquim Peiró}
\author[c]{David Moxey}

\cortext[author] {Corresponding author.\\\textit{E-mail address:} h.wustenberg@imperial.ac.uk}
\address[a]{\rev{Department of Aeronautics, Imperial College London, Exhibition Road, London SW7 2AZ, United Kingdom}}
\address[b]{\rev{McLaren Racing Ltd., McLaren Technology Centre, Chertsey Road, Woking, Surrey GU21 4YH, United Kingdom}}
\address[c]{\rev{Department of Engineering, King's College London, Strand Campus, London WC2R 2LS, United Kingdom}}

\begin{abstract}
\rev{Scale-resolving simulations of high-Reynolds-number incompressible flow are often limited by the Courant--Friedrichs--Lewy (CFL) restriction of explicit time-stepping schemes, leading to small steps and long times-to-solution.}
\rev{We systematically compare two implicit formulations of the velocity correction scheme -- a linear-implicit approach and a sub-stepping (semi-Lagrangian) method -- with a standard semi-implicit formulation in a high-order spectral/hp element framework.}
The schemes are assessed in terms of stability limits, temporal accuracy, and computational performance for implicit large-eddy simulation of the Imperial Front Wing benchmark, a complex high Reynolds number geometry with curved surfaces that imposes strict CFL constraints.
\rev{The sub-stepping scheme extends the stable time step size by a factor of 20, and the linear-implicit scheme by up to a factor of 100 when equal-order velocity and pressure spaces are used.
While increasing the cost per time step, the linear-implicit scheme reduces the time-to-solution by up to a factor of 8.7 at this largest step, whereas the sub-stepping scheme reduces it by only 12\%.}
\rev{With application targets of 1\% for mean lift and 5\% for mean drag, the fastest accepted linear-implicit case reduces time-to-solution by a factor of 1.53 at ten times the CFL-limited step. The main-plane separation bubble is already substantially shortened at this step, showing that force agreement can mask changes in transition-sensitive flow structure.}
The results quantify the trade-off between stability, accuracy, and computational cost for implicit velocity correction schemes on complex geometries and provide guidance for selecting time integration strategies in large-scale scale-resolving simulations.
\end{abstract}
\end{frontmatter}

\section{Introduction}
\label{sec.introduction}
Scale-resolving simulations are required to accurately capture unsteady flow phenomena such as vortex shedding, separation, and laminar–turbulent transition. 
However, their widespread use remains limited by the high computational cost associated with advancing the governing equations over physically relevant time intervals, particularly for high Reynolds number flows around complex geometries.

The computational effort is governed by the time-to-solution, defined as
\begin{equation*}
 \tsol=\ndt \times \tdt,
\end{equation*}
where $\ndt$ is the number of time steps and $\tdt$ the cost per step. 
In this study, we assume a fixed spatial discretisation to isolate the influence of the time-stepping scheme on both quantities\rev{, with one stated exception for the largest linear-implicit time steps}.
Explicit schemes incur low per-step cost but are constrained by the Courant–Friedrichs–Lewy (CFL) condition, which enforces small time step sizes and therefore large $\ndt$. 
Implicit schemes relax this restriction and allow larger time steps, but increase $\tdt$. 
Moreover, enlarging the time step may degrade temporal accuracy. 
Selecting an appropriate time integration strategy therefore requires balancing stability, accuracy, and computational efficiency.

We investigate this trade-off for incompressible flow around a complex high Reynolds number geometry with curved surfaces that impose restrictive CFL limits. 
The study is conducted within the class of velocity correction schemes \cite{Karniadakis1991,Guermond2003}, which split the incompressible Navier–Stokes equations into a pressure Poisson problem and Helmholtz problems for the velocity components. 
These schemes form the basis of several large-scale incompressible flow solvers, including Neko \cite{Jansson2024}, ExaDG \cite{arndt_exadg_2020,Fehn2021}, Nek5000 \cite{Merzari2020}, NekRS \cite{Fischer2022}, and \nektarpp \cite{Cantwell2015,Moxey2020}.

Most implementations employ a semi-implicit formulation in which diffusion is treated implicitly and advection explicitly. 
While computationally efficient, this formulation remains subject to a CFL restriction arising from the explicit advection term, which can become severe for high-order discretisations and geometrically complex domains.

Several strategies have been proposed to relax this restriction.
\begin{itemize}
    \item Fully implicit formulations treat advection implicitly, resulting in nonlinear systems. Baek et al. \cite{Baek2011} reported stability improvements of up to two orders of magnitude in time step size using fixed-point iterations with Aitken relaxation, but with substantially increased per-step cost.
    \item Operator-integration-factor-splitting (OIFS) methods \cite{Maday1990,Couzy1995,Deville2002} address the restriction by sub-stepping the advection equation in a pseudo-time. Sherwin and co-workers \cite{Sherwin2003} applied this approach within a velocity correction framework, enabling larger outer time steps while introducing additional cost proportional to the number of sub-steps. Similarly, semi-Lagrangian formulations \cite{Xiu2001,Xiu2005} permit substantial increases in time step size by tracing characteristics, but have primarily been investigated for laminar flows.
    \item Linear-implicit approaches \cite{Simo1994,Dong2010} avoid nonlinear solves by linearising the advection operator through a single Picard iteration, extending stability limits at the expense of time-dependent system matrices. Energy-stable formulations \cite{Lin2019,Lin2020} further enhance stability properties but require additional Helmholtz and Poisson solves per time step, effectively doubling the linear solver workload.
\end{itemize}

In this work, we focus on the sub-stepping and linear-implicit formulations, which aim to relax the CFL restriction without incurring the full cost of nonlinear implicit solvers. 
In contrast to prior studies that mainly examined canonical or laminar configurations, we conduct a systematic comparison of these schemes for implicit large-eddy simulation (LES) of a complex high Reynolds number geometry using a unified high-order spectral/hp element discretisation. 
By holding the spatial approximation fixed\rev{, apart from the pressure polynomial order of the linear-implicit scheme at $\dt \geq 20\,\dtcfl$,} we isolate the effect of temporal integration and directly quantify (i) achievable stability limits across the initial transient, (ii) accuracy degradation as a function of time step size, and (iii) time-to-solution per convective time unit in the quasi-steady regime.
To our knowledge, this is the first comprehensive evaluation of implicit velocity correction strategies under realistic wall-resolved LES (WRLES) conditions on complex geometry, providing quantitative guidance for selecting time integration schemes in large-scale incompressible flow solvers.
\rev{A complementary comparison of the same three schemes on a well-established canonical benchmark, the flow past a circular cylinder, is provided in a companion paper \cite{Liosi2026}.}

For turbulent simulations, two performance regimes are relevant. 
Statistical quantities are obtained by sampling the quasi-steady turbulent state until convergence of target measures, such as drag. 
However, simulations must first traverse an initial transient before reaching this regime. 
This motivates a separate performance objective: reducing the time required to reach the end of the transient.
During this phase, less restrictive accuracy targets may permit larger stable steps, provided the solution is allowed to re-equilibrate after the step is reduced and before statistics are gathered.
Once sampling begins, accuracy requirements for the target observables limit this flexibility.
The present cost measurements are made in the quasi-steady regime; timing the complete transient-to-sampling workflow remains future work.

As a representative benchmark, we consider the Imperial Front Wing (IFW) geometry \cite{Buscariolo2019}, a complex multi-component wing configuration characterised by curved surfaces and high Reynolds number flow. 
The full geometry has been studied experimentally \cite{Pegrum2006} and numerically using LES \cite{Lombard2017,Buscariolo2022}. 
We focus on an extruded slice of the geometry (eIFW), which preserves the essential flow physics while reducing computational complexity.
A WRLES benchmark data set for the eIFW is available in the work of Slaughter et al.~\cite{Slaughter2023}, and recent work by Ntoukas et al.~\cite{Ntoukas2025} compared explicit and implicit LES approaches for this configuration.

The time-stepping schemes are implemented within the open-source spectral/hp element framework \nektarpp \cite{Cantwell2015,Moxey2020}, which enables high-order discretisations on curved geometries and scalable parallel execution \cite{Lindblad2024}.
The low numerical dissipation of the high-order formulation allows the present study to employ an implicit LES (iLES) strategy, in which subgrid-scale effects are represented by the discretisation without introducing an explicit turbulence model.

The remainder of the paper is organised as follows. Section~\ref{sec.methodology} introduces the semi-implicit, \rev{sub-stepping, and linear-implicit} velocity correction schemes and summarises the spectral/hp element discretisation for WRLES.
Section~\ref{sec.benchmark} describes the benchmark configuration and computational setup.
Section~\ref{sec.verification} verifies spatial resolution and statistical convergence.
Sections~\ref{sec.stability}–\ref{sec.performance} present stability limits, accuracy assessment, and computational performance comparisons.
Finally, conclusions are drawn in Section~\ref{sec.conclusions}\rev{; the appendices discuss the iterative solver tolerance, the Reynolds-number difference to the reference, and the probe locations}.

\section{Methodology}
\label{sec.methodology}

We consider the incompressible Navier-Stokes equations
\begin{align}
    \frac{\partial \mathbf{u}}{\partial t} + \mathbf{u} \cdot \nabla \mathbf{u} + \nabla p - \nu \nabla^2 \mathbf{u} &= \mathbf{f} \label{eq.navierStokes}, \\
    \nabla \cdot \mathbf{u} &= 0 \label{eq.incompressibility}, \\
    \mathbf{u} \rvert_{\Gamma_D} &= \mathbf{u}_D \label{eq.velocityDirichlet}, \\
    \frac{\partial \mathbf{u}}{\partial \mathbf{n}}\Big\rvert_{\Gamma_N} &= \mathbf{u}_N \label{eq.velocityNeumann}.
\end{align}
They form a nonlinear system of equations which couple the $d$-dimensional velocity field $\mathbf{u}$ with an auxiliary pressure variable $p$.
We assume unit density $\rho=1$ and denote the kinematic viscosity by $\nu$.
Assuming an open bounded domain $\Omega$ with boundary $\Gamma = \partial \Omega$, the equations are supplemented with Dirichlet boundary conditions $\mathbf{u}_D$ for the inflow and walls with or without slip.
We also use Neumann boundary conditions $\mathbf{u}_N$ for the outflow and farfield boundaries.

\subsection{Time-stepping with velocity correction schemes}
\label{sec.timestepping}
\begin{figure}[htbp]
\centering
     \begin{subfigure}[b]{0.3\textwidth}
        \centering
        \includegraphics[width=\textwidth]{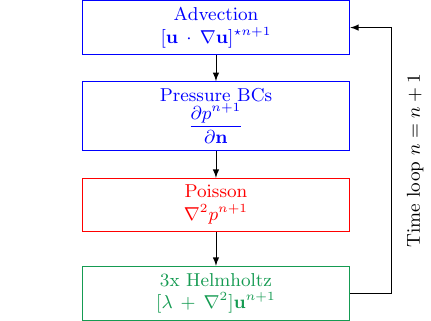}
        \caption{Semi-implicit scheme.}
        \label{fig.vcsAlgorithmSemi}
    \end{subfigure}
    \hfill
    \begin{subfigure}[b]{0.3\textwidth}
        \centering
        \includegraphics[width=\textwidth]{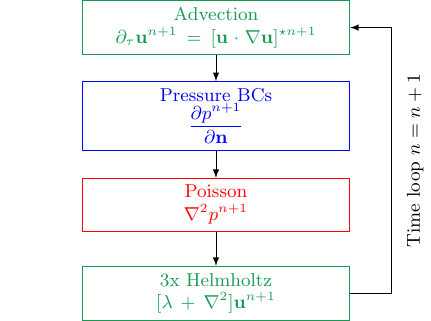}
        \caption{Sub-stepping scheme.}
        \label{fig.vcsAlgorithmSub}
    \end{subfigure}
    \hfill
    \begin{subfigure}[b]{0.3\textwidth}
        \centering
        \includegraphics[width=\textwidth]{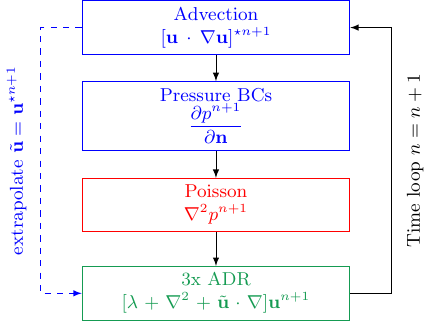}
        \caption{Linear-implicit scheme.}
        \label{fig.vcsAlgorithmLinear}
    \end{subfigure}
    \caption{The algorithm of the velocity correction scheme in semi-implicit, 
    sub-stepping and linear-implicit formulation. Note that all schemes are linear 
    within a time step and only the sub-stepping scheme uses sub-iterations for the 
    advection term \rev{which is an explicit evaluation for the semi-implicit and 
    linear-implicit formulation. The linear-implicit scheme additionally constructs 
    an extrapolated velocity field $\tilde{\mathbf{u}}$ for linearising the implicit 
    advection operator within the Advection-Diffusion-Reaction (ADR) problem.}}
    \label{fig.vcsAlgorithm}
\end{figure}

Time integration is performed using velocity correction splitting schemes \cite{Guermond2003}, which decouple pressure and velocity solves and form the backbone of many large-scale incompressible flow solvers. 
The splitting separates the incompressible Navier–Stokes equations into a pressure Poisson problem and $d$ velocity equations per time step, as illustrated in figure \ref{fig.vcsAlgorithm}. 
We first introduce the semi-implicit formulation and subsequently describe the modifications required for the sub-stepping and linear-implicit schemes.

The overview in figure \ref{fig.vcsAlgorithm} shows the significant steps of the velocity correction algorithm and compares the semi-implicit, sub-stepping and linear-implicit time-stepping schemes.
We begin by introducing the semi-implicit algorithm depicted in figure \ref{fig.vcsAlgorithmSemi} along with the formal definition, and, next, highlight differences in the sub-stepping and linear-implicit algorithms.

\subsubsection{Semi-implicit formulation}
Figure \ref{fig.vcsAlgorithmSemi} depicts the semi-implicit algorithm.
The semi-implicit formulation evaluates the advection term explicitly at the beginning, extrapolates it to $J_e$-th order and updates the pressure boundary condition $\frac{\partial p^{n+1}}{\mathbf{\partial n}}$.
Next, it solves the pressure Poisson problem to update the pressure to the new time $p^{n+1}$.
Finally, the $d$ velocity Helmholtz problems advance each velocity component to $\mathbf{u}^{n+1}$ with implicit diffusion treatment.

The velocity correction scheme uses an implicit-explicit (IMEX) approach with explicit treatment of the nonlinear term and implicit treatment of the diffusion term.
We employ a $J_i$th order backwards difference formula (BDF) for the time discretisation using the notation
\begin{align}
    \hat{\mathbf{u}} = \sum_{q=0}^{J_i-1} \alpha_q \mathbf{u}^{n-q} = \left\{\begin{array}{llr}
        \mathbf{u}^n, & \gamma=1, & \text{if } J_i=1,\\
        2\mathbf{u}^n - \frac{1}{2}\mathbf{u}^{n-1}, & \gamma=\frac{3}{2}, & \text{if } J_i=2.
    \end{array}\right.
\end{align}
We extrapolate explicit terms with stiffly-stable \rev{extrapolation} coefficients of order $J_e$ as
\begin{align}\label{eq.extrapolationBetaCoefficients}
    \mathbf{g}^{\star n+1} = \sum_{q=0}^{J_e-1} \beta_q \mathbf{g}^{n-q} = \left\{\begin{array}{lr}
        \mathbf{g}^{n} &\text{if } J_e=1,\\
        2 \mathbf{g}^{n} - \mathbf{g}^{n-1} &\text{if } J_e=2,
    \end{array}\right.
\end{align}
where $\mathbf{g}$ is a placeholder for an arbitrary term, for example, the nonlinear term.

\rev{Unless otherwise stated, all results presented in this paper use $J_i = J_e = 2$, i.e.\ second-order backward differentiation (BDF2) for the implicit terms combined with second-order stiffly-stable extrapolation of the explicit terms. The semi-implicit and linear-implicit schemes are confirmed second-order accurate in time using the method of manufactured solutions in \cite{Wuestenberg2025Thesis}; the sub-stepping scheme's temporal accuracy is discussed separately below.}

The velocity correction scheme requires solving a pressure Poisson problem of the form
\begin{align}
    \nabla^2 p^{n+1} &= \nabla \cdot \left[ -\frac{1}{\dt} (\gamma \mathbf{u}^{\dagger} - \hat{\mathbf{u}}) - [\mathbf{u} \cdot \nabla \mathbf{u}]^{\star n+1} - \nu \nabla \times \nabla \times \mathbf{u}^{\star n+1} + \mathbf{f}^{n+1} \right], \label{eq.pressureSubstep} \\
    \nabla \cdot \mathbf{u}^{\dagger} &= 0, \label{eq.incompressibilityAssumption} \\
    \mathbf{n} \cdot \mathbf{u}^{\dagger} \rvert_{\Gamma_D} &= \mathbf{n} \cdot \mathbf{u}_D^{n+1}. \label{eq.pressureSubstepDirichlet}
\end{align}
\rev{Here $\mathbf{u}^{\dagger}$ denotes the divergence-free intermediate velocity of the pressure step, and $\mathbf{f}$ a body force.
Karniadakis et al.~\cite{Karniadakis1991} showed that consistent pressure boundary conditions are essential for higher-order temporal accuracy; they are obtained by writing the explicitly treated diffusion term in the rotational form, which accounts for incompressibility \cite{Guermond2006}.}
The consistent pressure boundary condition is
\begin{equation}
    \frac{\partial p}{\partial \mathbf{n}}^{n+1} 
    = 
    \mathbf{n} \cdot \left[ 
    - \frac{\gamma}{\dt} \mathbf{u}_D^{n+1} 
    - [\mathbf{u} \cdot \nabla \mathbf{u}]^{\star n+1} 
    - \nu \nabla \times \nabla \times \mathbf{u}^{\star n+1} 
    + \mathbf{f}^{n+1} \right]
    .\label{eq.pressureBoundaryCondition}
\end{equation}

The second step of the algorithm is computing the new velocity components $\mathbf{u}^{n+1} = [u^{n+1}, v^{n+1}, w^{n+1}]$, for $d = 3$ dimensions.
The semi-implicit formulation solves a Helmholtz problem of the form
\begin{align}
    \frac{\gamma}{\dt} \mathbf{u}^{n+1} 
    - \nu \nabla^2 \mathbf{u}^{n+1} 
    &= 
    \frac{1}{\dt} \hat{\mathbf{u}} 
    - [\mathbf{u} \cdot \nabla \mathbf{u}]^{\star n+1} 
    - \nabla p^{n+1}
    + \mathbf{f}^{n+1} 
    , \label{eq.velocityHelmholtz} \\
    \mathbf{u}^{n+1}|_{\Gamma_D} &= \mathbf{u}_D^{n+1}.
\end{align}
Notably, the Helmholtz problem leads to a symmetric and positive definite matrix with constant coefficients when we discretise in space.
This reduces computational overhead compared to the linear-implicit velocity operator introduced below.

\subsubsection{The sub-stepping scheme}
\label{sec.substepping-scheme}
The sub-stepping scheme \cite{Sherwin2003} relaxes the CFL restriction of the semi-implicit formulation by resolving the advection term through multiple pseudo-time substeps within each physical time step.
Instead of evaluating the nonlinear term explicitly at the beginning of the step, the material derivative is advanced in a Lagrangian frame, while the pressure and viscous terms remain in the Eulerian frame.

Starting from the material derivative
\begin{equation}
    \frac{D \mathbf{u}}{Dt}
    = \frac{\partial \mathbf{u}}{\partial t}
    + \mathbf{u} \cdot \nabla \mathbf{u},
\end{equation}
the convective transport is computed by solving an auxiliary unsteady advection equation in pseudo-time $\tau$,
\begin{equation}\label{eq.substepping}
    \frac{\partial \hat{\mathbf{u}}}{\partial \tau}
    + \mathbf{u} \cdot \nabla \hat{\mathbf{u}} = 0,
\end{equation}
over $\ndtau$ substeps within a single physical time step.
\rev{In the present simulations the pseudo-time advection uses a second-order Runge--Kutta (improved Euler) scheme, following \cite{Sherwin2003}, and $\ndtau$ is chosen at every time step such that the substep CFL number does not exceed $0.5$.}

The resulting velocity field provides a stabilised evaluation of the advection term, which is then used in the pressure Poisson and velocity Helmholtz problems.
\rev{Compared to the semi-implicit formulation, the sub-stepping scheme thus replaces the explicit extrapolation of the nonlinear term with a pseudo-time advection solve; the pressure Poisson and velocity Helmholtz operators are unchanged, and only their right-hand sides differ (see figure~\ref{fig.vcsAlgorithm}).}
The improved stability is obtained at the expense of additional computational work, which scales proportionally with the number of pseudo-time substeps $\ndtau$ executed within each physical time step.

\rev{Splitting the advective transport from the implicit pressure/diffusion solve is known to introduce an $\mathcal{O}(\Delta t)$ error that is independent of the pseudo-time integration order, first reported for this class of scheme in reference \cite{Couzy1995} and confirmed by \cite{Sherwin2003} on an analytic Kovasznay-flow test.
Repeating the manufactured-solution study of \cite{Wuestenberg2025Thesis} for the sub-stepping scheme reproduces this behaviour: whereas the semi-implicit and linear-implicit schemes converge at the expected second order, the sub-stepping scheme converges at only first order across the tested range of time step sizes.
The splitting error is problem-dependent; at $\dt=\dtcfl$ the force predictions of all three schemes agree closely (section~\ref{sec.accuracy-scheme}), so it is not significant there.
Because it grows linearly with $\dt$, it contributes to the sub-stepping errors at larger time step sizes reported in section~\ref{sec.accuracy-dt}, where it cannot be separated from other temporal errors.}

\subsubsection{The linear-implicit formulation}
\label{sec.linearimplicit-formulation}
The linear-implicit velocity correction scheme relaxes the CFL restriction by linearising the advection operator within the velocity step.
\rev{Instead of reusing the explicit non-linear term, as in the semi-implicit formulation,} the advection velocity at time level $n+1$ is approximated as
\begin{equation}
    \mathbf{u}^{n+1} \cdot \nabla \mathbf{u}^{n+1}
    \approx
    \tilde{\mathbf{u}} \cdot \nabla \mathbf{u}^{n+1}, \label{eq.linearAdvectionApprox}
\end{equation}
where the advection velocity $\tilde{\mathbf{u}}$ is obtained by extrapolation from previous time levels \rev{as proposed in the work \cite{Simo1994}}.
\rev{Note that the extrapolation of the advection velocity, defined as $\tilde{\mathbf{u}} = \mathbf{u}^{\star n+1}$ following equation \eqref{eq.extrapolationBetaCoefficients}, is evaluated as part of the explicit advection step, see figure \ref{fig.vcsAlgorithmLinear} for reference}.
In this work, only this extrapolated form is considered.

This modification affects only the velocity equation, while the pressure Poisson problem \rev{has the same form as} in the semi-implicit formulation with explicit evaluation of the nonlinear term.
\rev{In the present implementation, however, its right-hand side is assembled in weak form, as an inner product against the gradient of the pressure test functions, whereas the semi-implicit and sub-stepping schemes use the strong, pointwise divergence of the extrapolated velocity field \cite{Wuestenberg2025Thesis}; section~\ref{sec.stability} returns to this difference.}
The resulting velocity system becomes an advection–diffusion–reaction (ADR) problem,
\begin{align}\label{eq.velocityADR}
    \frac{\gamma}{\dt} \mathbf{u}^{n+1}
    - \nu \nabla^2 \mathbf{u}^{n+1}
    + \tilde{\mathbf{u}} \cdot \nabla \mathbf{u}^{n+1}
    =
    \frac{1}{\dt} \hat{\mathbf{u}}
    - \nabla p^{n+1}
    + \mathbf{f}^{n+1}.
\end{align}

Compared to the semi-implicit scheme, the linear-implicit formulation therefore replaces the symmetric Helmholtz operator with a non-symmetric ADR operator containing time-dependent advection coefficients.
This enhances stability without requiring nonlinear iterations\rev{, at the cost of assembling and storing a non-symmetric matrix at every time step (section~\ref{sec.linear-solvers})}.

\subsection{Spatial discretisation via spectral/hp elements}
We use the spectral/hp element method to discretise in space.
Spectral/hp elements enable a hierarchical approach to higher-order finite element methods \cite{Sherwin1997}.
It builds basis functions with a modal structure such that the initial modes are linear and higher modes are additive as higher-order polynomials \cite{Dubiner1991}.
\rev{Within each element, the solution is expanded in the $N_m$ basis functions $\phi_{pqr}$ as}

\begin{equation}
    u(x_i,y_j,z_k) = \sum^{N_{m}}_{pqr} \phi_{pqr}(x_i,y_j,z_k) \rev{\hat{u}_{pqr}},
\end{equation}

where \rev{$\hat{u}_{pqr}$} are the coefficients of the basis functions.
The above summation shows that any evaluation of the solution is typically an order $\mathcal{O}(N^6)$ operation in three dimensions, assuming the number of modes is equivalent to the number of degrees of freedom within an element $N_m \approx N_\text{DoF}$.
\rev{The modal basis is, however, amenable to sum factorisation on all element types, including simplices, which reduces this cost to $\mathcal{O}(N^4)$ operations \cite{Karniadakis2013}.}

Consequently, the dominant computational cost per time step arises from the solution of the global linear systems associated with the Poisson, Helmholtz, or ADR operators.
Any modification of the time-stepping scheme that introduces additional solves or requires reconstruction of system matrices therefore directly impacts the overall time-to-solution.

\subsection{Stabilisation of aliasing errors}
\label{sec.svv}
Wall-resolved LES does not fully resolve the turbulent energy spectrum, which can lead to numerical instabilities due to unresolved scales and aliasing errors. 
We adopt the iLES approach \cite{Sagaut2006}, where spatial resolution determines the separation between resolved and unresolved scales.

To prevent unbounded energy growth, we employ spectral vanishing viscosity (SVV) \cite{Tadmor1989}. 
Specifically, we use the discontinuous Galerkin (DG)-mimicking SVV kernel proposed by Moura et al.~\cite{Moura2020}, which mitigates wave trapping in regions of rapidly varying mesh resolution and has demonstrated robustness for complex geometries \cite{Mengaldo2021}\rev{; the SVV diffusion coefficient is set to $0.4$}.

Aliasing errors also arise from nonlinear advection and geometric mapping of curved elements \cite{Mengaldo2015,Ferrer2017}. 
To control these effects, we apply global over-integration using quadrature rules accurate for cubic nonlinearities. 
Following \cite{Mengaldo2015}, the number of quadrature points $Q$ for polynomial order $P$ has to satisfy
\begin{equation}
    Q \geq P + \frac{P_\text{order}}{2} + \frac{3}{2},
\end{equation}
where $P_\text{order}$ denotes the order of nonlinearity, including contributions from geometric projection.
\rev{For the present velocity order $P_v = 4$ we use $Q = 7$ quadrature points per direction ($6$ in the collapsed Gauss--Radau directions).}

\subsection{Linear system solvers}
\label{sec.linear-solvers}
The dominant cost per time step arises from the solution of the linear systems associated with the Poisson, Helmholtz, and ADR problems in \eqref{eq.pressureSubstep}, \eqref{eq.velocityHelmholtz}, and \eqref{eq.velocityADR}.
We formulate these problems in weak form, leading to discrete systems $A x = b$ after taking $L^2$ inner products with the basis functions $\phi$.

We apply static condensation (substructuring) \cite{guyan_reduction_1965,irons_structural_1965}, reducing the global system to boundary degrees of freedom while interior modes are eliminated locally.
This lowers both the rank and condition number of the condensed system and significantly reduces iterative solver cost \cite{Karniadakis2013}.

We summarise the linear system solvers for each equation and time-stepping scheme in table \ref{tab.solverSetup}.
The linear systems for the Poisson and Helmholtz problems are symmetric and positive definite and we solve them with a conjugate gradient (CG) method \cite{hestenes_methods_1952}.
The ADR matrix, however, is a general matrix due to the advection coefficients and, hence, we employ the restarted generalised minimum residual (GMRES) method \cite{Saad2003} for iterative solution.
The preconditioner for the pressure Poisson problem is the inverse diagonal of the matrix known as Jacobi preconditioner.
The Helmholtz matrix uses the low-energy block preconditioner introduced in reference \cite{Sherwin2001}.
This preconditioner further transforms the Schur complement to a low-energy space which reduces the coupling between modes of the basis functions $\phi$.
The low-energy block preconditioner is equally applied to the ADR matrix.
The stopping criterion for all iterative solvers is \rev{an absolute residual tolerance} $|r| < \rev{\epsilon}$ where $|r| = |b - A x|$ is the magnitude of the residual and $|\cdot|$ indicates the L2 norm\rev{, with $\epsilon = 10^{-4}$ for the pressure and $\epsilon = 10^{-2}$ for the velocity systems (table~\ref{tab.solverSetup})}.

\begin{table}
    \begin{tabular}{lllll}
        Equation & Scheme & Iterative solver & Preconditioner & Tolerance \\
        \toprule
        \textcolor{blue}{Advection} & Sub-stepping & - & - & - \\
        \midrule
        \textcolor{red}{Pressure} & All & CG & Diagonal & $1\times10^{-4}$ \\
        \midrule
        \textcolor{ForestGreen}{Velocity} & Semi-implicit & CG & LowEnergy & $1\times10^{-2}$ \\
        \textcolor{ForestGreen}{Velocity} & Sub-stepping & CG & LowEnergy & $1\times10^{-2}$ \\
        \textcolor{ForestGreen}{Velocity} & Linear-implicit  & GMRES & LowEnergy & $1\times10^{-2}$ \\
        \bottomrule
    \end{tabular}
    \caption{The configuration of linear solvers for each equation and time-stepping scheme. We also show the preconditioner and tolerance for the stopping criterion. Note colours correspond to the algorithm overviews in figure \ref{fig.vcsAlgorithm}.}
    \label{tab.solverSetup}
\end{table}

A key difference between the CFL-restricted semi-implicit scheme and the sub-stepping and linear-implicit schemes is the additional work incurred per physical time step. 
For the sub-stepping scheme, an auxiliary advection problem is solved over $\ndtau$ pseudo-time substeps. 
Even with inexpensive substeps, this additional work limits the attainable speed-up.

For the linear-implicit scheme, the advection-diffusion-reaction matrix contains time-dependent coefficients.
Consequently, the statically condensed system must be reconstructed at every time step, increasing assembly and solver overhead.
When time step sizes remain close to those of the semi-implicit scheme, this reconstruction cost may offset stability gains.

\subsection{Parallelisation and mesh decomposition}
Large-scale simulations are performed using distributed-memory parallelisation based on the message passing interface (MPI).
The dominant computational cost arises from the solution of global linear systems, making efficient domain decomposition and communication minimisation essential for scalability.
The computational mesh is partitioned across MPI ranks using a nested dissection algorithm provided by the Scotch graph partitioning library \cite{pellegrini_scotch_2012,chevalier_pt-scotch_2008}, which minimises communication between subdomains.

Within each MPI rank, operator evaluations are optimised through vectorised implementations using single instruction multiple data (SIMD) instructions via advanced vector extensions (AVX). To further improve computational efficiency, elements are locally grouped according to their polynomial order and geometric type, enabling the selection of specialised operator kernels and improving data locality \cite{Cantwell2011}.

Parallel input and output operations are implemented using the hierarchical data format (HDF5), allowing scalable reading of large meshes and writing of solution fields during runtime.

A detailed description of the parallelisation strategy and performance optimisations implemented in \nektarpp can be found in \cite{Lindblad2024}.

\section{An industrial benchmark configuration}
\label{sec.benchmark}

\subsection{Geometry and flow configuration}

We consider the extruded Imperial Front Wing (eIFW) configuration derived from the Imperial Front Wing geometry \cite{Buscariolo2019}.
The geometry consists of a multi-element wing operating in ground effect and exhibits laminar–turbulent transition and strong pressure gradients along the wing elements.
These characteristics make the configuration a demanding benchmark for wall-resolved large-eddy simulation.

\begin{figure}[htbp]
    \centering
    \includegraphics[width=\textwidth]{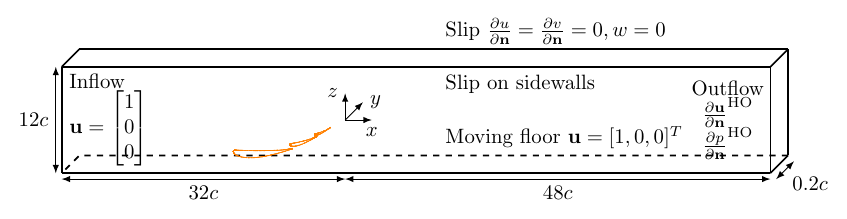}
    \caption{Computational domain and boundary conditions with wing elements. Note that the wing elements are enlarged by factor $4$ for visibility.}
    \label{fig.eifw-domain}
\end{figure}

The computational domain is obtained by extracting a slice of the geometry at $Y=250$~mm and extruding it in the spanwise direction over a length $L_y=0.05$~m.
The resulting configuration contains the three primary aerodynamic components of the front wing: the main plane, first flap, and second flap, see figure~\ref{fig.eifw-domain}.
The ride height is set to \rev{$h/c = 0.36$}, consistent with previous experimental and numerical investigations of the full geometry \cite{Pegrum2006,Lombard2017,Buscariolo2022}.

\subsection{Computational domain and boundary conditions}

Figure~\ref{fig.eifw-domain} shows the computational domain and boundary conditions.
The streamwise domain length is $60c$, where $c=0.25$~m denotes the chord of the main plane.
The domain height is $12c$ and the spanwise width is $0.2c$, following the spanwise sensitivity analysis reported in \cite{Slaughter2023}.
The resulting blockage ratio is approximately $4.3\%$, comparable to previous studies.

A uniform inflow velocity is prescribed together with a moving floor to mimic the relative motion between the vehicle and the road.
Slip boundary conditions are applied at the lateral boundaries, similar to \cite{Ntoukas2025,OSullivan2025}.
At the outlet we employ the robust energy-stable boundary condition introduced in \cite{Dong2014} to prevent instabilities caused by vortical structures exiting the computational domain\rev{, which allows the domain to be truncated downstream}.

\begin{figure}[htbp]
    \centering
    \includegraphics[width=0.80\textwidth]{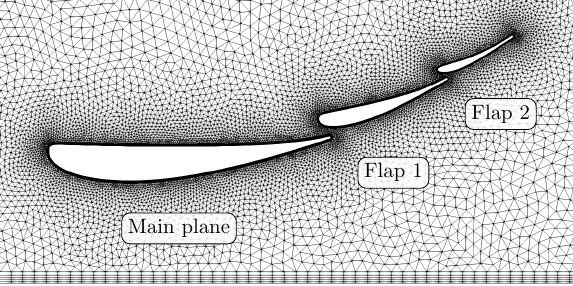}
    \caption{A cross-section of the near field mesh highlighting the fine boundary layer mesh on the wing elements and floor.}
    \label{fig.eifw-mesh-near}
\end{figure}

\subsection{Flow parameters and nondimensionalisation}

The simulations are performed in nondimensional form using a reference velocity $U=1$.
The Reynolds number is defined using the chord length of the main plane as
\begin{equation}
    \mathrm{Re} = \frac{Uc}{\nu}.
\end{equation}
We consider \rev{$\mathrm{Re} = 1.73 \times 10^5$}, corresponding to a kinematic viscosity $\nu = c/Re$.
This value is approximately \rev{$21\%$} lower than that used in previous studies \cite{Slaughter2023,Ntoukas2025}\rev{. The difference results from an error in the simulation setup, and repeating the simulation campaign was not feasible because of its computational cost. Instead, \ref{ap.reynolds} quantifies its effect: the dominant frequencies of the lift spectrum are insensitive to it, while the mean forces and the transition location change moderately.}

Physical time is reported in convective time units (CTU), defined as
\begin{equation}
    t^* = \frac{tU}{c},
\end{equation}
which corresponds to the time required for the flow to travel one chord length.

\subsection{Pre-processing and initial conditions}
\label{sec.preprocessing}

The computational mesh is generated using the NekMesh preprocessing tools within \nektarpp \cite{Kirilov2024}.
A linear finite element mesh is first created from the CAD geometry using StarCCM+ (version 2021.1) \cite{Starccm}.
This mesh is subsequently elevated to a high-order representation and curved using a variational optimisation procedure to accurately capture the geometric surfaces \cite{Green2024}.

Initial conditions are obtained from a Reynolds-averaged Navier–Stokes (RANS) solution computed with a $k$–$\epsilon$ turbulence model in StarCCM+.
The resulting solution is interpolated onto the high-order mesh to provide a physically consistent starting point for the wall-resolved LES.
A detailed description of this workflow within \nektarpp is given in \cite{Lindblad2024}.

The resulting high-order mesh and boundary-layer discretisation are described in the following subsection.

\subsection{Computational mesh}
\label{sec.mesh}

The mesh is designed to accurately resolve the near-wall flow while limiting resolution in the far wake where detailed comparison with reference data is not required.
Prism layers are employed along the wing elements and moving floor to capture boundary-layer development and transition, see figure~\ref{fig.eifw-mesh-near}.
Additional refinement is applied near the leading and trailing edges where strong pressure gradients occur.

The final mesh consists of $418\times10^3$ prismatic elements in the boundary-layer regions and $668\times10^3$ tetrahedral elements in the outer flow.
A Taylor–Hood discretisation is used with polynomial order $P_v=4$ for velocity and $P_p=3$ for pressure, resulting in $21.1\times10^6$ global degrees of freedom and $54.8\times10^6$ local degrees of freedom.
\rev{The linear-implicit simulations at $\dt \geq 20\,\dtcfl$ instead use an equal-order discretisation with $P_p = P_v = 4$, see section~\ref{sec.stability}. The quadrature is unchanged, but the number of local pressure coefficients increases from approximately $30.1\times10^6$ to $54.8\times10^6$.}

\subsection{Boundary layer resolution}
\label{sec.boundary_layer_resolution}

The near-wall mesh resolution is assessed using wall units $\Delta x^+$, $\Delta y^+$ and $\Delta z^+$ defined using the friction velocity $u_\tau$,
\begin{equation}
\Delta x^+ = \frac{u_\tau}{\nu}\Delta x,
\qquad
u_\tau = \sqrt{\tau_w}.
\end{equation}

\begin{figure}[htbp]
     \centering
    \includegraphics[width=\textwidth]{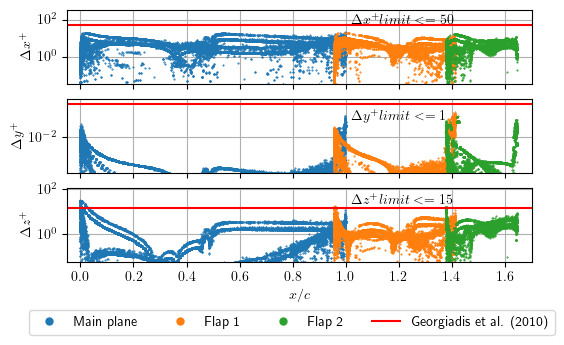}
    \caption{Distribution of wall units $\Delta x^+, \Delta y^+, \Delta z^+$ for the time-averaged flow field on all wing elements. Note that $\Delta y^+$ is the wall-normal direction \rev{of the local wall frame}.}
    \label{fig.verification-wallunit}
\end{figure}

Figure~\ref{fig.verification-wallunit} shows the distribution of wall units along the wing surfaces.
\rev{The wall units are evaluated in a local frame on each surface, with $\Delta x^+$ along the surface in the streamwise direction, $\Delta y^+$ normal to the wall and $\Delta z^+$ spanwise; this differs from the global coordinates, in which the span is along $y$ and the vertical along $z$.}
The streamwise and wall-normal spacings remain well below the commonly accepted limits for wall-resolved LES \cite{Georgiadis2010}.
The spanwise spacing $\Delta z^+$ also satisfies these limits over most of the surface.
Slightly larger values occur near the leading edges, where the flow remains laminar and the resolution requirements are less restrictive.

Overall, the eIFW configuration represents a demanding benchmark for wall-resolved LES due to the combination of curved multi-element geometry, laminar–turbulent transition, and high Reynolds number flow.
The fine near-wall resolution together with the geometric complexity results in restrictive CFL limits for explicit time integration, making this configuration particularly suitable for evaluating the stability and performance of implicit time-stepping schemes.

\section{Verification of the numerical setup}
\label{sec.verification}

To verify the spatial and temporal resolution of the numerical setup introduced in section \ref{sec.benchmark}, we compare the present simulations with previously published wall-resolved LES results for the extruded IFW configuration.
In particular, we use the reference dataset of Slaughter et al. \cite{Slaughter2023}, which provides detailed information on mean flow features, transition mechanisms, and spectral characteristics of the flow.

Verification is performed using the semi-implicit time-stepping scheme with a time step size $\Delta t = 1\times10^{-5}$, which satisfies the CFL restriction and serves as the baseline configuration for subsequent comparisons of implicit time-stepping schemes\rev{; section~\ref{sec.stability} shows that the semi-implicit scheme is unstable at twice this time step size}.
In the following, we use the shorthand
\begin{equation}\label{eq.dtcfl}
    \dtcfl = 1\times10^{-5}
\end{equation}
for this reference time step size.

\subsection{Convergence of statistics}
\label{sec.convergence}

The Marginal Standard Error Rule (MSER) \cite{White1997} is used to estimate the end of the initial transient phase.
Following the work in Bergmann et al.~\cite{Bergmann2021}, MSER provides a robust criterion based on minimising the variance of truncated time series for large-eddy simulations.

For the reference simulation, MSER applied to the lift and drag coefficients suggests the end of the initial transient at approximately $T_0 \approx 7$\ctu.
However, to ensure conservative statistical sampling and consistency with previous studies \cite{Slaughter2023}, we select $T_0 = 30$\ctu for all subsequent analysis.

\begin{comment}
According to the MSER, for a timeseries of size $N$ where each sample is expreseed by $g_i$, the timestamp $d$ of the timeseries, that signifies the end of the initial transient is defined as

\begin{equation}
   d = min\left[ \frac{1}{(N-d)^2} \sum_{i=d}^{N-1}(g_i - \overline{g}_{N,d}) \right]
\end{equation}

where $\overline{g}_{N,d}$ is the truncated average of the timeseries with starting index $d$ until the end of the timeseries. This non-linear equation is solved iteratively until the timestamp that minimises the mean square error of the truncated timeseries is found.
\end{comment}

%

%

\begin{figure}[htbp]
    \centering
    \begin{subfigure}[b]{0.45\textwidth}
        \centering
        \includegraphics[width=\textwidth]{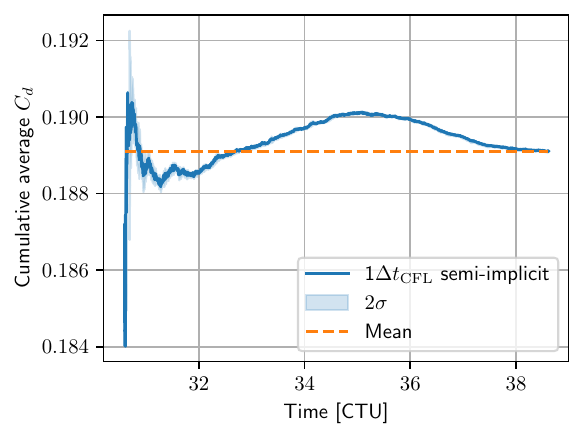}
        \caption{Drag coefficient.}
    \end{subfigure}
    \hfill
    \begin{subfigure}[b]{0.45\textwidth}
        \centering
        \includegraphics[width=\textwidth]{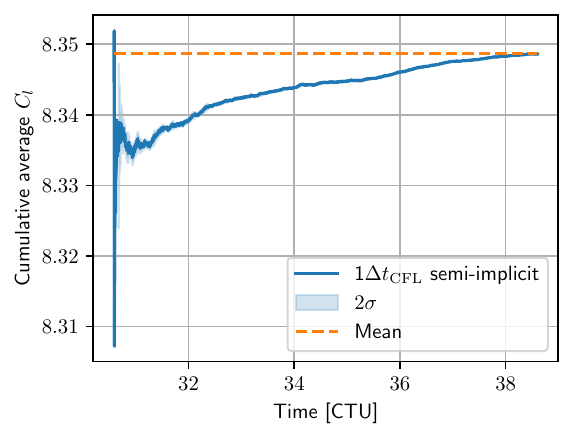}
        \caption{Lift coefficient.}
    \end{subfigure}
    \caption{Cumulative average for the lift and drag coefficients with averaging window $T \in [30, 38]$.}
    \label{fig.ca_forces}
\end{figure}

\begin{figure}[htbp]
    \centering
    \begin{subfigure}[b]{0.45\textwidth}
        \centering
        \includegraphics[width=\textwidth]{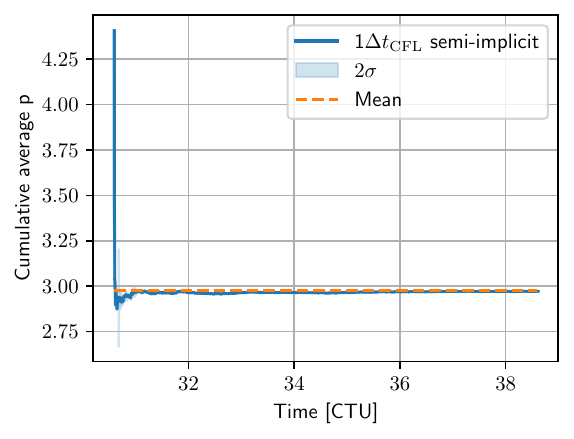}
        \caption{Pressure $p$.}
        \label{fig.ca_probe8_p}
    \end{subfigure}
    \hfill
    \begin{subfigure}[b]{0.45\textwidth}
        \centering
        \includegraphics[width=\textwidth]{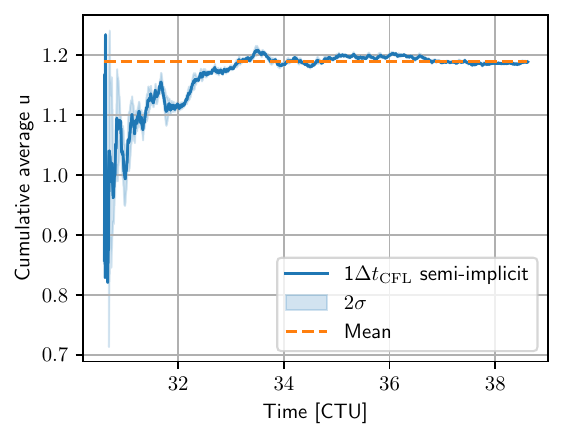}
        \caption{Streamwise velocity $u$.}
        \label{fig.ca_probe8_u}
    \end{subfigure}
    \caption{Cumulative average and rolling standard deviation for the pressure and streamwise velocity.}
    \label{fig.ca_probe8}
\end{figure}

The duration of the averaging window is $8$\ctu, consistent with \cite{Slaughter2023}.
\rev{All cases are averaged over an $8$\ctu window starting at $T \geq 30$\ctu, except for the linear-implicit scheme at $\dtcfl$, for which the available window is $T \in [20.2, 28.2]$\ctu; this window still starts well after the end of the initial transient identified by MSER ($T_0 \approx 7$\ctu).}
Statistical convergence is assessed using cumulative averages and fluctuations about the mean for the global force coefficients in figure \ref{fig.ca_forces} and for a pointwise probe.
We present probe data for the main-plane location \#8, see table~\ref{tab.probe_locations} for coordinates, which exhibits the largest standard deviation and, hence, provides the most stringent test for statistical convergence.
Figure \ref{fig.ca_probe8} shows the cumulative averages for the pressure $p$ and streamwise velocity $u$.
Both pointwise and global signals show high variations within the initial $1$\ctu and stable convergence towards the mean within the total $8$\ctu window.
For all monitored quantities, the cumulative mean varies by less than $1\%$ over the final $20\%$ of the sampling window, indicating adequate convergence of the statistics.

The established statistical convergence ensures that both time-averaged quantities and spectral estimates are not contaminated by initial transient effects.

\subsection{Comparison of global quantities}

Table~\ref{tab.eifw-mean-coefficients-literature} compares time-averaged lift and drag coefficients with values reported in the literature.
The force coefficients are defined as
\begin{equation}
    C(F) = \frac{F}{\frac{1}{2} \rho U^2 A},
\end{equation}
where $F$ denotes either the lift force \rev{$F_z$} or drag force $F_x$, $U$ is the freestream velocity, $\rho$ the density, and $A$ the reference surface area.

The predicted mean lift coefficient $\overline{C}_l$ agrees closely with the reference study of Slaughter et al.~\cite{Slaughter2023} and lies within the range reported by other LES studies.
The drag coefficient $\overline{C}_d$ is \rev{about $11\%$} higher than in \cite{Slaughter2023} but comparable to the explicit LES results reported by Ntoukas et al.~\cite{Ntoukas2025}.
Overall, the force coefficients indicate that the global aerodynamic behaviour of the flow is captured correctly.

These integral quantities provide a first check of the global aerodynamic response, while more sensitive transition diagnostics are examined in the following subsections.

\begin{table}[htb]
    \centering
    \caption{Time-averaged force coefficients and dominant PSD peaks reported in the literature and obtained in the present study for the extruded IFW configuration.}
    \label{tab.eifw-mean-coefficients-literature}
    \begin{tabular}{llll}
        Source & $\overline{C}_l$ & $\overline{C}_d$ & PSD peaks in $\mathrm{St}$ \\
        \toprule
        Slaughter et al. \cite{Slaughter2023}   & -8.33     & 0.17      & $30, 40, 60, 140, 200$ \\
        Ntoukas et al. \cite{Ntoukas2025} iLES  & -8.6821   & 0.1725    & $23, 33, 50$ \\
        Ntoukas et al. \cite{Ntoukas2025} eLES  & -8.5674   & 0.1815    & $23, 33, 50$ \\
        O'Sullivan et al. \cite{OSullivan2025}  & -8.212    & 0.177     & $25, 38, 53, 115, 158$ \\
        \midrule
        Current study                           & -8.3486   & 0.1891   & $21, 29, 40$ \\
        \bottomrule
    \end{tabular}
\end{table}

\subsection{Mean flow features}
\label{sec.meanflow}

\begin{figure}[htbp]
    \centering
    \begin{subfigure}[b]{\textwidth}
        \centering
        \includegraphics[width=\textwidth]{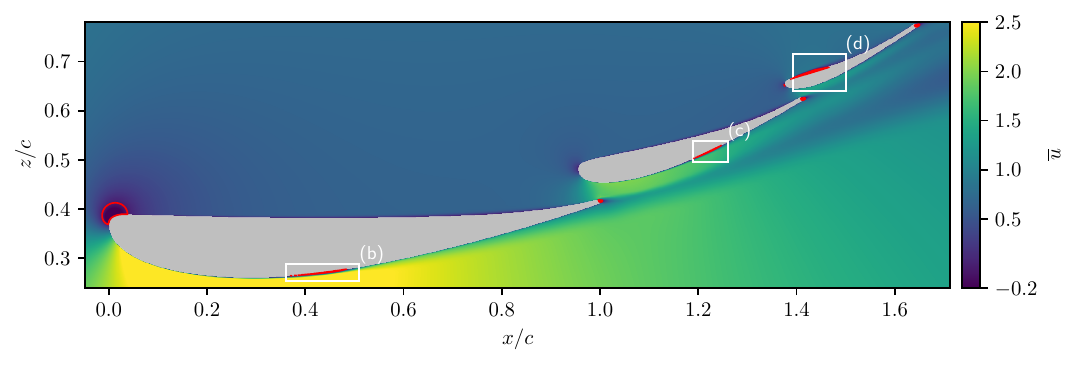}
        \caption{\rev{Mean streamwise velocity $\overline{u}$ with the contour $\overline{u}=0$ in red, which also marks the leading-edge stagnation points and trailing-edge recirculation regions; the boxes mark (b)--(d).}}
        \label{fig.verification-velmag-overview}
    \end{subfigure}
    \\[1ex]
    \begin{subfigure}[b]{\textwidth}
        \centering
        \includegraphics[width=\textwidth]{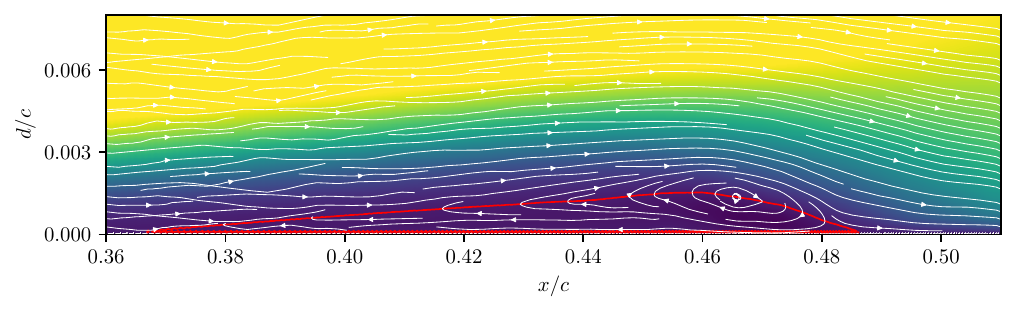}
        \caption{\rev{Main plane, suction side.}}
        \label{fig.verification-velmag-main}
    \end{subfigure}
    \\[1ex]
    \begin{subfigure}[b]{0.48\textwidth}
        \centering
        \includegraphics[width=\textwidth]{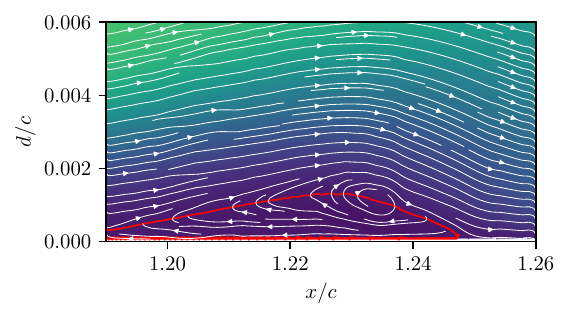}
        \caption{\rev{Flap~1, suction side.}}
        \label{fig.verification-velmag-flap1}
    \end{subfigure}
    \hfill
    \begin{subfigure}[b]{0.48\textwidth}
        \centering
        \includegraphics[width=\textwidth]{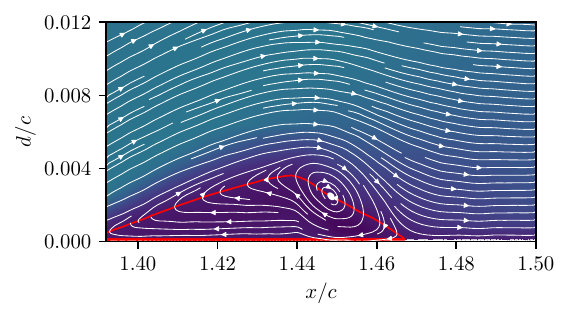}
        \caption{\rev{Flap~2, pressure side.}}
        \label{fig.verification-velmag-flap2}
    \end{subfigure}
    \caption{\rev{Time-averaged flow on the mid-span plane ($y=-0.1c$) for the semi-implicit scheme at $\dtcfl$. Panels (b)--(d) use the colour scale of (a) and show mean streamlines (white) over the separation bubbles; the ordinate $d$ is the vertical distance from the wall, stretched relative to $x/c$.}}
    \label{fig.verification-velmag}
\end{figure}

Figure~\ref{fig.verification-velmag}\rev{(a)} shows the time-averaged \rev{streamwise velocity $\overline{u}$ on the mid-span plane}.
Contours of zero streamwise velocity identify recirculation regions associated with laminar separation bubbles (LSBs).
\rev{The close-ups in figure~\ref{fig.verification-velmag}(b)--(d) resolve these regions with mean streamlines. Because the bubbles are boundary-layer thin, with heights below $0.4\%$ of the chord, they are plotted against the vertical distance from the wall, $d$, stretched relative to $x$.}

The predicted flow topology agrees well with previous studies \cite{Slaughter2023,Ntoukas2025}.
Laminar separation bubbles are observed on the suction sides of the main plane and first flap, while a separation region forms on the pressure side of the second flap.
\rev{At mid-span, the main-plane bubble extends from $x/c\approx0.37$ to $0.48$, the flap~1 bubble from $x/c\approx1.19$ to $1.25$, and the flap~2 bubble from $x/c\approx1.39$ to $1.47$. In each case the mean streamlines leave the wall, enclose a closed recirculation, and reattach downstream.
The contour $\overline{u}=0$ also marks the stagnation points at the leading edges and small recirculation regions at the trailing edges of the elements.}
The \rev{locations} of these structures \rev{agree with} the reference simulations\rev{, although the main-plane bubble is longer, see below}, indicating that the present setup captures the dominant transition mechanisms of the boundary-layer flow.
\rev{The three-dimensional organisation of these separation bubbles, and its sensitivity to the time-stepping scheme and time step size, is examined in sections~\ref{sec.bubble-organisation-scheme} and~\ref{sec.bubble-organisation-dt}.}

\subsection{Pressure distribution and transition to turbulence}

\begin{figure}[htbp]
    \centering
    \begin{subfigure}[b]{\textwidth}
        \centering
        \includegraphics[width=\textwidth]{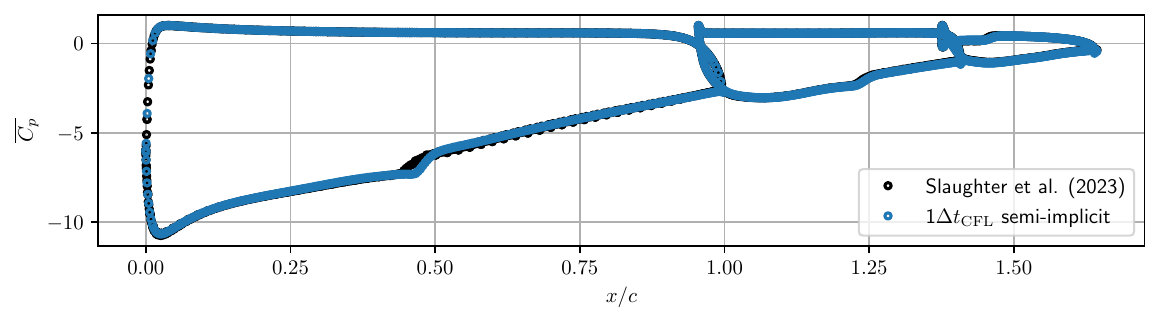}
        \caption{Pressure coefficient $\overline{C}_p$.}
        \label{fig.verification-cp}
    \end{subfigure}
    \newline
    \begin{subfigure}[b]{\textwidth}
        \centering
        \includegraphics[width=\textwidth]{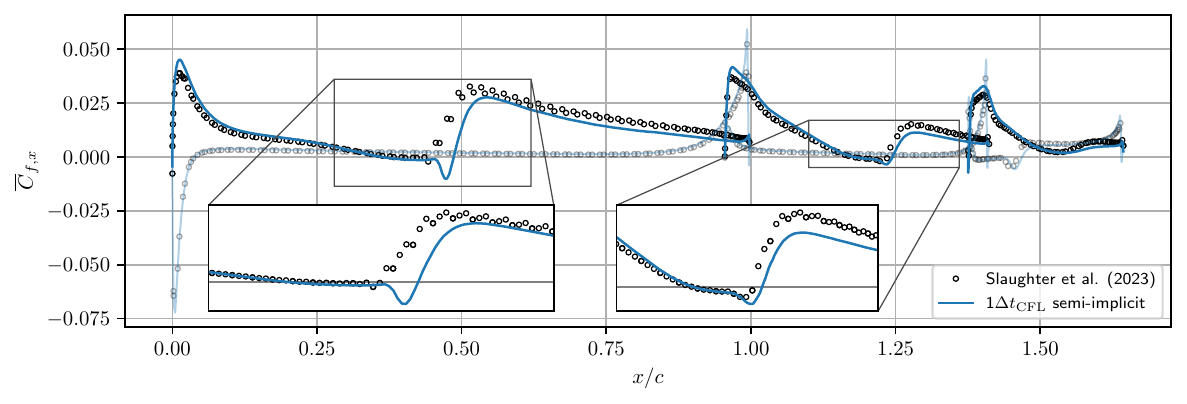}
        \caption{\rev{Streamwise skin-friction coefficient $\overline{C}_{f,x}$. Insets: suction side of the main plane and flap~1.}}
        \label{fig.verification-cf}
    \end{subfigure}
    \caption{Time-averaged pressure and skin-friction coefficients for the three wing elements. \rev{In (b), the suction side is drawn in full colour and the pressure side is shaded lighter; the insets enlarge the suction-side transition regions (marked by the boxes).}}
    \label{fig.verification-cp-cf}
\end{figure}

The surface pressure coefficient $\overline{C}_p$ and skin-friction coefficient $\overline{C}_f$ are shown in figure~\ref{fig.verification-cp-cf}.
\rev{The skin friction is shown as its signed streamwise component, $\overline{C}_{f,x} = 2\,\overline{\tau}_{w,x}/(\rho U^2)$, so that regions of reversed near-wall flow ($\overline{C}_{f,x}<0$) and the separation and reattachment points ($\overline{C}_{f,x}=0$) can be read off directly. Because only the $x$-component is shown, $\overline{C}_{f,x}$ tends to zero where the surface is nearly vertical, i.e.\ at the leading edges.}
The pressure distribution along all three wing elements agrees well with the reference data, indicating that the global pressure loading is accurately captured.

The skin-friction coefficient highlights the laminar–turbulent transition along the surfaces.
For the main plane, the transition location is in good agreement with the reference data, while the reattachment point is slightly shifted downstream relative to \cite{Slaughter2023}\rev{: in the span-averaged skin friction, the suction-side flow separates at $x/c \approx 0.36$ (reference: $0.38$) and reattaches at $x/c \approx 0.48$ (reference: $0.45$)}.
\rev{At the present Reynolds number, the laminar skin friction upstream of the main-plane bubble is $11$--$16\%$ above the reference, consistent with $\mathrm{Re}^{-1/2}$ scaling; at matched Reynolds number (\ref{ap.reynolds}) it agrees to within $2\%$.
The lower Reynolds number also accounts for about half of the longer main-plane bubble ($0.12c$ compared to $0.066c$; $0.091c$ at matched Reynolds number).
The remaining differences are consistent with higher numerical dissipation in the present simulation: the bubble is still longer, and downstream of reattachment the turbulent skin friction is $10$--$25\%$ below the reference (peak $\overline{C}_{f,x} = 0.028$ compared to $0.033$), a difference that does not decrease at matched Reynolds number.
The reference uses gradient-jump penalty stabilisation at polynomial order $6$ \cite{Slaughter2023}, whereas the present simulation uses the DG-kernel SVV at order $4$, which is known to be more diffusive at moderate polynomial orders \cite{Moura2022}.}

The transition locations on the first and second flap also agree closely with previously reported results, further supporting the adequacy of the near-wall resolution.
\rev{The suction-side separation bubble on flap~1 extends over $1.18 \lesssim x/c \lesssim 1.24$ (reference: $1.17$--$1.24$), and the pressure-side bubble on flap~2 over $1.39 \lesssim x/c \lesssim 1.47$ (reference: $1.38$--$1.46$). Overall, separation and reattachment on all three elements agree with the reference to within $0.01$--$0.03c$, so the flow topology is well reproduced.}

\subsection{Resolution of temporal scales}

To verify that the simulations resolve the relevant temporal scales of the flow, we analyse the frequency spectrum of the total lift coefficient.
The power spectral density (PSD) is estimated using Welch's method \cite{Welch1967}.

The lift signal is sampled \rev{$12\,500$ times per convective time unit (every second time step)} over a duration of 8 convective time units of statistically stationary flow.
\rev{The PSD is computed from Hann-windowed segments of $12\,500$ samples ($1$\ctu) with $50\%$ overlap, i.e.\ $15$ periodograms, giving a frequency resolution of $\Delta\mathrm{St} = 1$.}
This averaging procedure reduces statistical noise while maintaining sufficient frequency resolution.

\begin{figure}[htbp]
    \centering
    \includegraphics[width=0.7\textwidth]{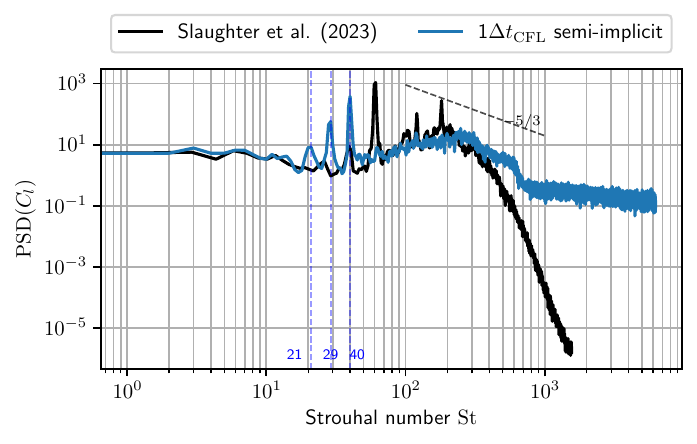}
    \caption{Power spectral density of the lift coefficient comparing semi-implicit prediction and reference data from \cite{Slaughter2023}.}
    \label{fig.verification-psd}
\end{figure}

Figure~\ref{fig.verification-psd} shows the PSD of the total lift coefficient compared with the reference results of Slaughter et al.~\cite{Slaughter2023}. The Strouhal number is defined as
\begin{equation}
    \mathrm{St} = \frac{fc}{U},
\end{equation}
where $f$ denotes the frequency.

The present simulations capture three dominant narrow-band peaks at $\mathrm{St} = \{21, 29, 40\}$.
These correspond to the low-frequency structures associated with the transition dynamics reported in the reference study.
Slaughter et al.~\cite{Slaughter2023} identified peaks at $\mathrm{St} = \{30, 40, 60\}$.
The shift \rev{of about $30\%$} towards lower frequencies in the present results is likely associated with the use of slip boundary conditions in the present setup.

Higher-frequency peaks reported at $\mathrm{St}=\{140, 200\}$ in \cite{Slaughter2023} are not observed here.
However, the overall spectral decay and dominant frequency band are consistent with previously reported LES results, indicating that the simulations adequately resolve the primary temporal dynamics of the flow.

The PSD shows a plateau in the energy decay for $\mathrm{St} \gtrsim 700$.
This behaviour is related to the tolerance of the iterative linear solvers used in the simulation, as discussed in \ref{ap.tolerance}.

\subsection{Turbulent kinetic energy}

\begin{figure}[htbp]
    \centering
    \includegraphics[width=\textwidth]{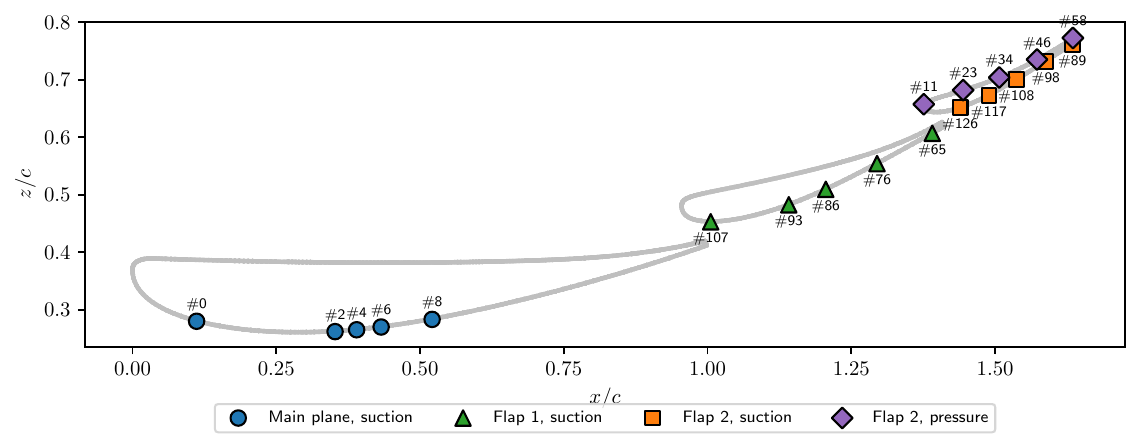}
    \caption{Mid-span section of the main plane, flap~1 and flap~2 (grey)
    with all point probes used in the analysis below. Point indices are
    those of the corresponding output files; exact coordinates are listed
    in table~\ref{tab.probe_locations}.}
    \label{fig.probe-geometry}
\end{figure}

Additional evidence for the transition mechanism is provided by the power spectral density of the turbulent kinetic energy (TKE) at \rev{five} probe locations along the suction side of the main plane.
\rev{The probes lie between $x/c \approx 0.11$ and $0.52$, from the laminar boundary layer to just downstream of the main-plane separation bubble ($0.36 \lesssim x/c \lesssim 0.48$),} and increasing probe index corresponds to increasing downstream distance from the leading edge.
Figure~\ref{fig.probe-geometry} shows these probes relative to the wing geometry, together with the further probes on flap~1 and flap~2 used later in the analysis, see section~\ref{sec.accuracy-scheme-spectra} and \ref{ap.probes}; exact coordinates for every probe are listed in table~\ref{tab.probe_locations}.

\begin{figure}[htbp]
    \centering
    \includegraphics[width=\textwidth]{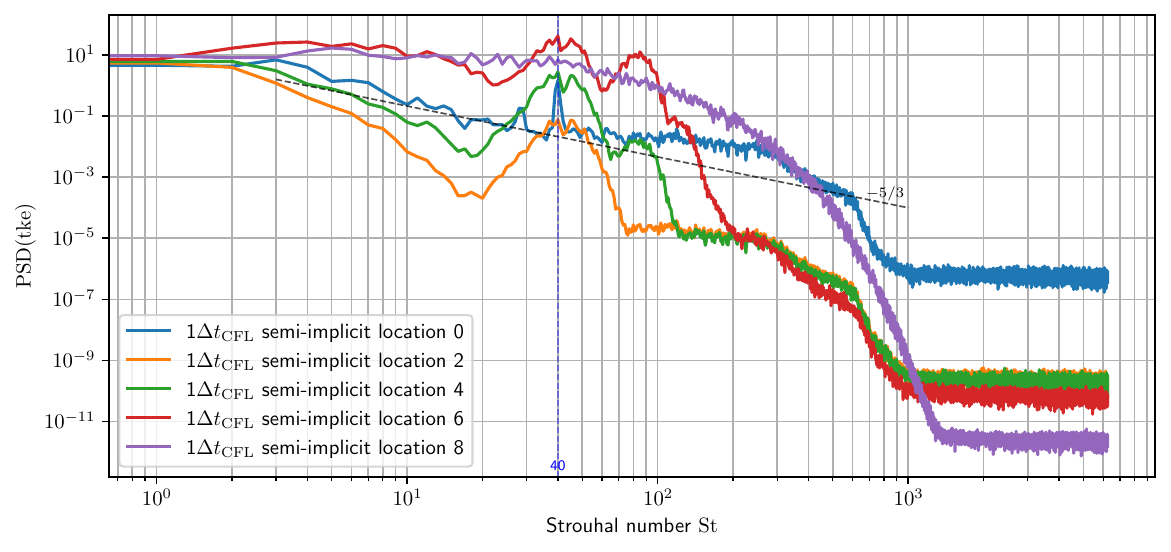}
    \caption{Power spectral density of turbulent kinetic energy at probe locations along the suction side of the main plane.}
    \label{fig.psd-his-mainplane-mid}
\end{figure}

Figure~\ref{fig.psd-his-mainplane-mid} shows that a narrow frequency band in the range $\mathrm{St} = 20$--$60$ is excited for most probe locations, except for the last two.
Moving \rev{downstream}, the TKE within this band increases, indicating amplification of disturbances associated with the laminar--turbulent transition on the main plane.

The most energetic frequency is found at approximately $\mathrm{St} = 40$, which is consistent with the travelling-wave instability reported in the literature for transition within the laminar separation bubble \cite{Slaughter2023}.
Slaughter et al.~\cite{Slaughter2023} reported a dominant peak at $\mathrm{St} = 60$ together with additional energetic content for $\mathrm{St} \geq 100$.
In contrast, the present spectra are more similar to those reported by \cite{Ntoukas2025,OSullivan2025}.
This difference may be related to the use of symmetry conditions in the spanwise direction in the present study, rather than the periodic treatment used in \cite{Slaughter2023}.

Overall, the comparisons of global force coefficients, mean flow structures, surface quantities, and spectral characteristics demonstrate good agreement with previously published LES studies.
These results confirm that the present spatial and temporal discretisation is sufficient to capture the key physical mechanisms of the extruded IFW flow, including laminar separation bubbles and transition to turbulence.
The configuration therefore provides a reliable baseline for evaluating the stability, accuracy and performance of different time-stepping schemes.

\section{Stability limits of the time-stepping schemes}
\label{sec.stability}

We assess the stability of the time-stepping schemes by determining the largest admissible time step size $\Delta t$ that yields a physically consistent solution.
\rev{A time step size is considered unstable if the simulation diverges.}
As a baseline, we consider the semi-implicit formulation and identify the maximum stable time step through a series of simulations.

The resulting reference time step in equation \eqref{eq.dtcfl} corresponds approximately to the stability limit imposed by the CFL restriction of the explicit advection term\rev{; in preliminary tests, the semi-implicit scheme was unstable at $\dt = 2\,\dtcfl$}.
The stability of larger time steps is evaluated by running each simulation from the initial condition ($t=0$\ctu) until $t=20$\ctu, which covers the transient phase during which the flow develops from the initial condition to a quasi-steady turbulent state.
This period is typically the most demanding for numerical stability due to the presence of energetic vortical structures.

To characterise the stability constraint imposed by the explicit advection operator, we evaluate the local CFL number using the estimator described in chapter 6 of \cite{Karniadakis2013}.
For spectral/hp element discretisations the elemental CFL number can be approximated as

\begin{equation}
    \text{CFL}^e \approx |V^{\text{st}}| \frac{c_\lambda P^2}{h},
\end{equation}

where $V^{\text{st}}$ denotes the local velocity magnitude within an element, $P$ is the polynomial order, $h$ is the characteristic element size, and $c_\lambda = 0.2$ is a scaling factor used for estimating the maximum eigenvalue of the advection operator.

 \begin{figure}[htbp]
    \centering
    \begin{tikzpicture}
        \node[anchor=south west,inner sep=0] (cflmain) at (0,0)
            {\includegraphics[width=\textwidth]{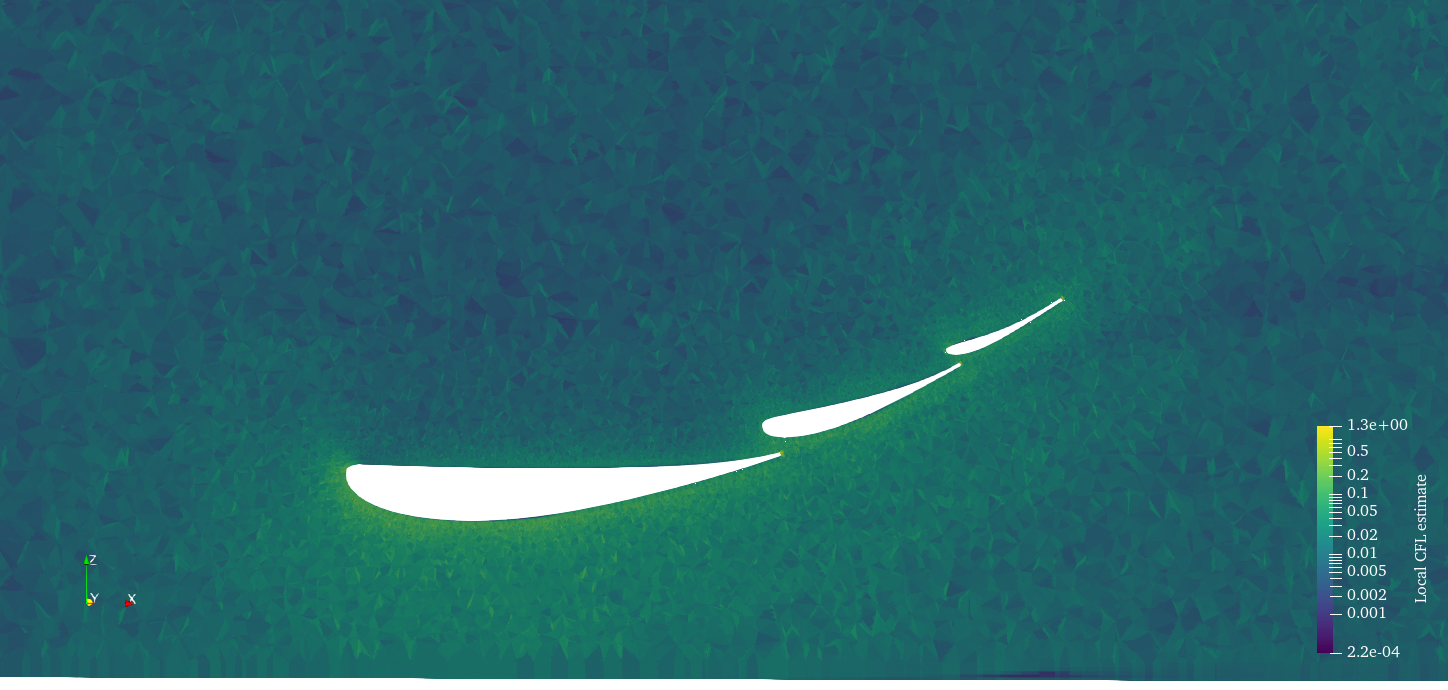}};
        \begin{scope}[x={($(cflmain.south east)-(cflmain.south west)$)},
                      y={($(cflmain.north west)-(cflmain.south west)$)}]
            \draw[red,thick] (0.4765,0.2735) rectangle ++(0.1550,0.1765);
            \coordinate (cflBoxCorner) at (0.4765,0.4500);
        \end{scope}
        \node[anchor=north west,inner sep=0,draw=red,very thick] (cflzoom) at
            ([xshift=2mm,yshift=-2mm]cflmain.north west)
            {\includegraphics[width=0.33\textwidth]{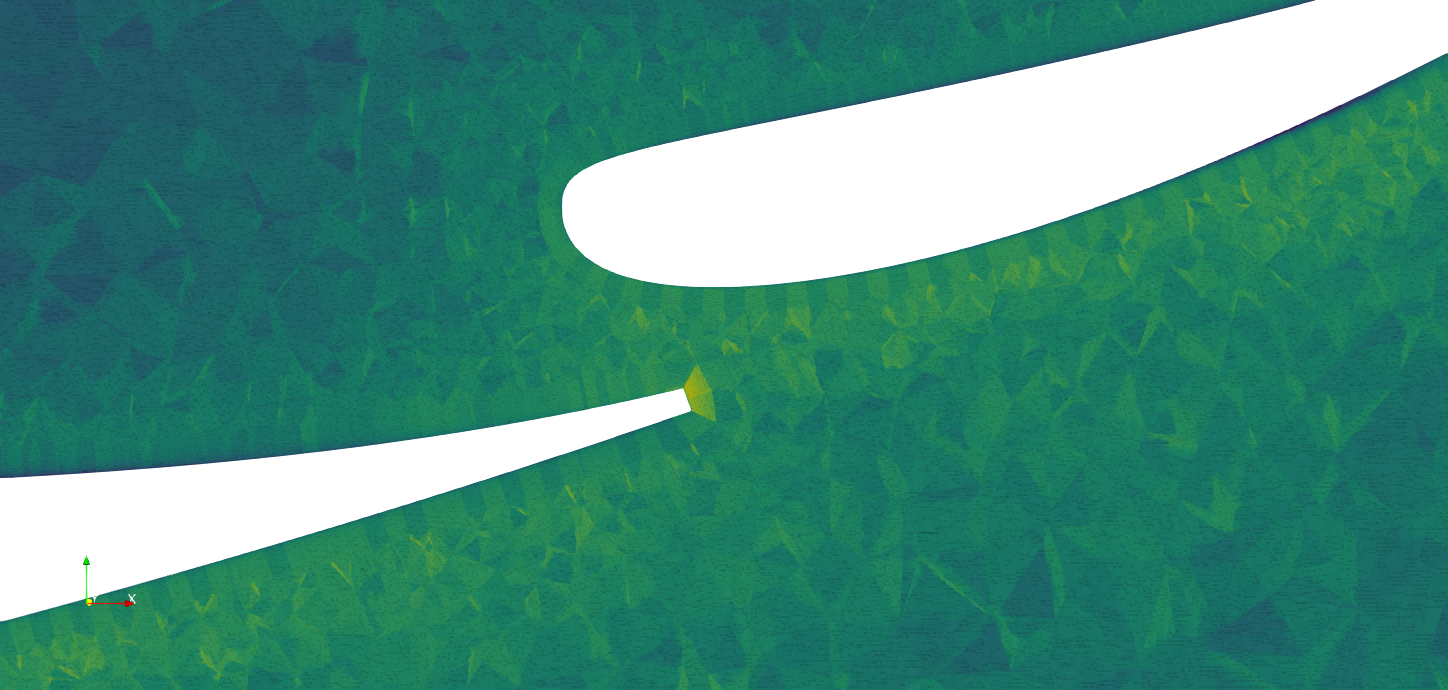}};
        \draw[red,thick] (cflzoom.south east) -- (cflBoxCorner);
    \end{tikzpicture}
    \caption{Time-averaged CFL estimate for the mid-plane and semi-implicit scheme with $\dt = \dtcfl$\rev{, shown on a logarithmic colour scale; the inset magnifies the main-plane trailing edge, where $\max(\mathrm{CFL}^e)$ occurs}.}
    \label{fig.cfl-space}
\end{figure}

Figure~\ref{fig.cfl-space} shows the spatial distribution of the time-averaged CFL estimate for the semi-implicit scheme using $\Delta t = \dtcfl$\rev{, on a logarithmic colour scale that resolves the strong variation between the freestream and the near-wall peaks}.
Over most of the computational domain the CFL number remains below $0.3$.
The largest values occur near the leading and trailing edges of the wing elements where the boundary-layer mesh is highly refined.

\rev{The maximum CFL value is observed at the trailing edge of the main plane (magnified inset), where the time-averaged value reaches approximately $\max(\mathrm{CFL}^e) \approx 1.3$.}
This region therefore determines the stability limit of the semi-implicit scheme.

\begin{figure}[htbp]
    \centering
    \includegraphics[width=0.70\textwidth]{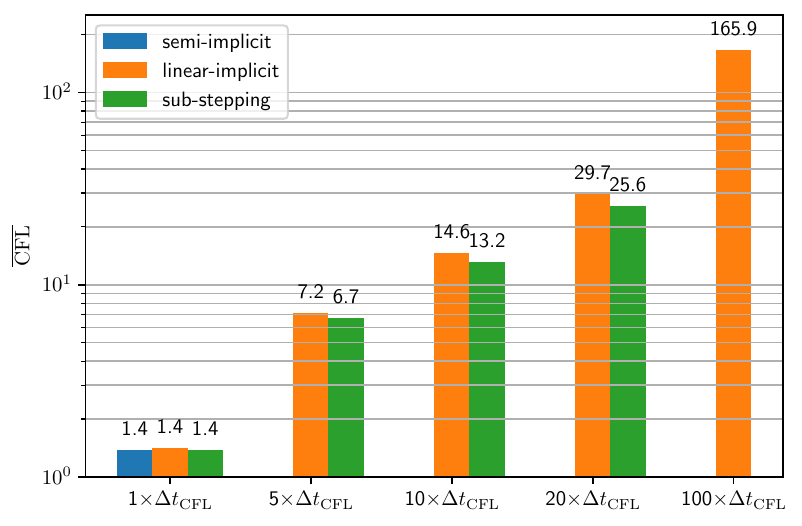}
    \caption{Time-average of the maximum local CFL estimate for all schemes at increasing time step size. Note values above each column are rounded to the first decimal. \eqordernote}
    \label{fig.cfl-scheme-dt}
\end{figure}

Figure~\ref{fig.cfl-scheme-dt} compares the maximum CFL number obtained for each time-stepping scheme as the time step size is increased.
Both the sub-stepping and linear-implicit formulations significantly extend the stability limit compared to the semi-implicit scheme.

For the sub-stepping scheme, stable simulations are obtained up to $\Delta t = 20\,\dtcfl$, corresponding to maximum CFL values of approximately \rev{$25.6$; in preliminary tests, it was unstable at $\dt = 30\,\dtcfl$}.
The increase in CFL scales nearly linearly with the time step size, indicating that the overall flow structures remain largely unchanged within this range.

\rev{With the Taylor--Hood spaces of section~\ref{sec.mesh}, the linear-implicit scheme was run stably up to $\Delta t = 10\,\dtcfl$.}

Interestingly, the linear-implicit scheme allows even larger time steps when equal-order approximation spaces are used for velocity and pressure ($P_v = P_p = 4$).
Under this discretisation, stable simulations were obtained up to $\Delta t = 100\,\dtcfl$\rev{, with a maximum CFL value of approximately $29.7$ at $20\,\dtcfl$.
All linear-implicit results at $\dt \geq 20\,\dtcfl$ presented in this paper, namely $20\,\dtcfl$ and $100\,\dtcfl$, therefore use equal-order spaces. Differences from the Taylor--Hood reference at these time steps combine the effect of the larger time step with that of the higher pressure polynomial order, and the two cannot be separated with the present data.}
In contrast, the semi-implicit and sub-stepping schemes did not exhibit improved stability under equal-order expansion.
\begin{revblock}
In all three schemes the velocity update is forced by the pressure gradient $-\nabla p^{n+1}$ from a Poisson problem of the same form, see equations \eqref{eq.velocityHelmholtz} and \eqref{eq.velocityADR}.
We offer two hypotheses for why only the linear-implicit scheme benefits from equal-order spaces.
\begin{itemize}
    \item \emph{Coercivity of the velocity operator.}
    The semi-implicit and sub-stepping schemes invert a symmetric, positive definite Helmholtz operator, so the response to a perturbation of the forcing is bounded by $\dt/\gamma$, independently of the flow.
    The advection--diffusion--reaction operator of the linear-implicit scheme satisfies the same bound only while the linearised advection term is skew-symmetric, which requires a divergence-free advection velocity $\tilde{\mathbf{u}}$.
    The extrapolated velocity is, however, only approximately divergence-free, and its divergence adds a symmetric contribution $-\tfrac{1}{2}\nabla\cdot\tilde{\mathbf{u}}$ that competes with the stabilising reaction term $\gamma/\dt$.
    As the time step size grows, the reaction term weakens and this contribution can make the operator lose its coercivity, so that perturbations of the forcing are no longer damped.
    Matching the pressure and velocity polynomial orders may improve the discrete divergence of the projected velocity, and hence of $\tilde{\mathbf{u}}$, which keeps the operator coercive up to larger time step sizes.
    \item \emph{Pressure boundary conditions.}
    In the semi-implicit and sub-stepping schemes the pressure boundary forcing is formed from the strong, pointwise divergence of the extrapolated velocity field, as suggested in \cite{Karniadakis1991}, so its accuracy is governed by the velocity polynomial order $P_v$ alone and is insensitive to $P_p$.
    In the linear-implicit scheme the pressure forcing is instead assembled in weak (variational) form, as an inner product against the gradient of the pressure test functions; the accuracy of this projection depends directly on $P_p$.
    Matching $P_p$ to $P_v$ therefore sharpens the pressure forcing only for the linear-implicit scheme.
    The work \cite{Wuestenberg2025Thesis} compares both pressure formulations directly.
\end{itemize}
We stress that both are hypotheses; a dedicated analysis of the discrete ADR operator would be required to confirm them.
\end{revblock}

Overall, both implicit formulations substantially relax the CFL restriction of the semi-implicit scheme, enabling time step sizes one to two orders of magnitude larger while maintaining numerical stability.

\section{Accuracy}
\label{sec.accuracy}

We assess temporal accuracy using both integral and pointwise quantities.
The integral quantities comprise the lift and drag forces and their spectral decomposition, which characterise the global aerodynamic response of the flow.
The pointwise quantities are the surface pressure and skin-friction coefficients, which are more sensitive to changes in the laminar–turbulent transition and boundary-layer development.

The analysis is divided into two parts.
First, we compare the semi-implicit, sub-stepping, and linear-implicit schemes at the reference time step size $\Delta t = \dtcfl$ in order to isolate differences caused solely by the time integration method.
Second, we examine the effect of increasing the time step size for the implicit schemes to quantify the trade-off between stability and temporal accuracy.

\subsection{Influence of the time-stepping scheme}\label{sec.accuracy-scheme}

\rev{We first compare the three schemes at the fixed time step size $\dt = \dtcfl$, using the time-averaged forces and surface data, the spectra of time-resolved signals, and the instantaneous and three-dimensional mean flow.}

\subsubsection{Time-averaged force coefficients}

\begin{figure}[htbp]
     \centering
     \begin{subfigure}[b]{0.45\textwidth}
        \centering
        \includegraphics[width=\textwidth]{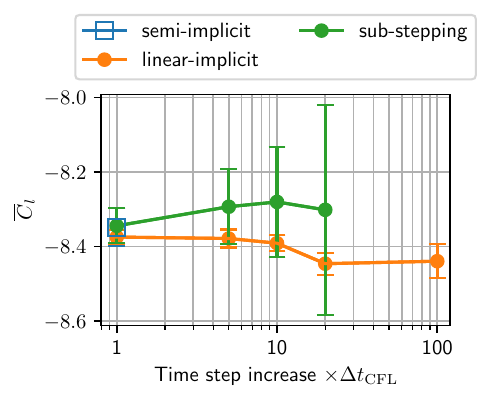}
        \caption{Time-averaged lift coefficient $\overline{C}_l$.}
        \label{fig.mean-force-coefficients-total-lift}
    \end{subfigure}
    \hfill
    \begin{subfigure}[b]{0.45\textwidth}
        \centering
        \includegraphics[width=\textwidth]{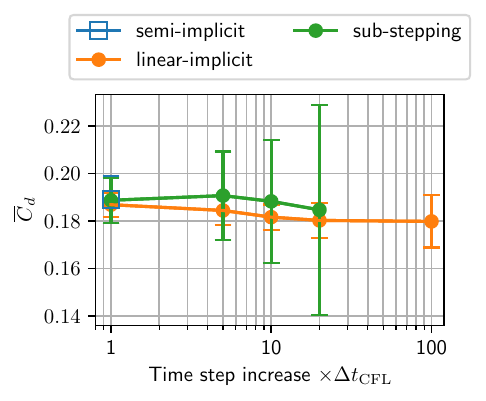}
        \caption{Time-averaged drag coefficient $\overline{C}_d$.}
        \label{fig.mean-force-coefficients-total-drag}
    \end{subfigure}
    \caption{The time-averaged lift coefficient $\overline{C}_l$ and drag coefficient $\overline{C}_d$ comparing the influence of the time-stepping scheme and increasing the time step size $\dt \geq \dtcfl$. The error bars show the standard deviation. \eqordernote}
    \label{fig.mean-force-coefficients-total}
\end{figure}

Figure~\ref{fig.mean-force-coefficients-total} compares the time-averaged lift and drag coefficients for the three time-stepping schemes at the reference time step size $\Delta t = \dtcfl$.

The sub-stepping and linear-implicit schemes predict force coefficients that are very close to the semi-implicit reference.
For the sub-stepping scheme, the mean lift differs by \rev{$0.05\%$}, while the mean drag differs by \rev{$0.21\%$}.
In comparison, the linear-implicit lift differs by $0.31\%$ and mean drag by \rev{$1.19\%$}.
These differences are small and indicate that, at the reference time step size, all schemes produce nearly equivalent global flow predictions.

\subsubsection{Time-averaged surface coefficients}
\label{sec.accuracy-scheme-surface}

The surface pressure and skin-friction coefficients provide a more sensitive measure of temporal accuracy than the integral forces, as they directly reflect the location of separation, transition, and reattachment along the wing elements.

\begin{figure}[!tb]
    \centering
    \begin{subfigure}[b]{\textwidth}
        \centering
        \includegraphics[width=\textwidth]{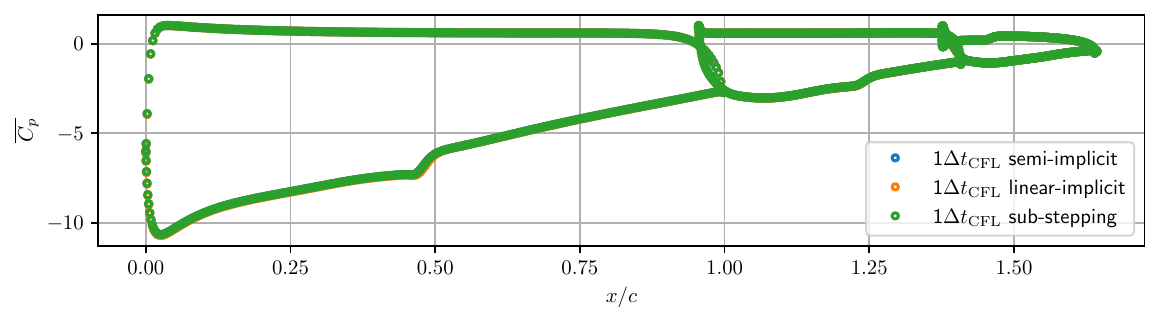}
        \caption{Pressure coefficient $\overline{C}_p$.}
        \label{fig.surface-pressure-scheme}
    \end{subfigure}
    \newline
    \begin{subfigure}[b]{\textwidth}
        \centering
        \includegraphics[width=\textwidth]{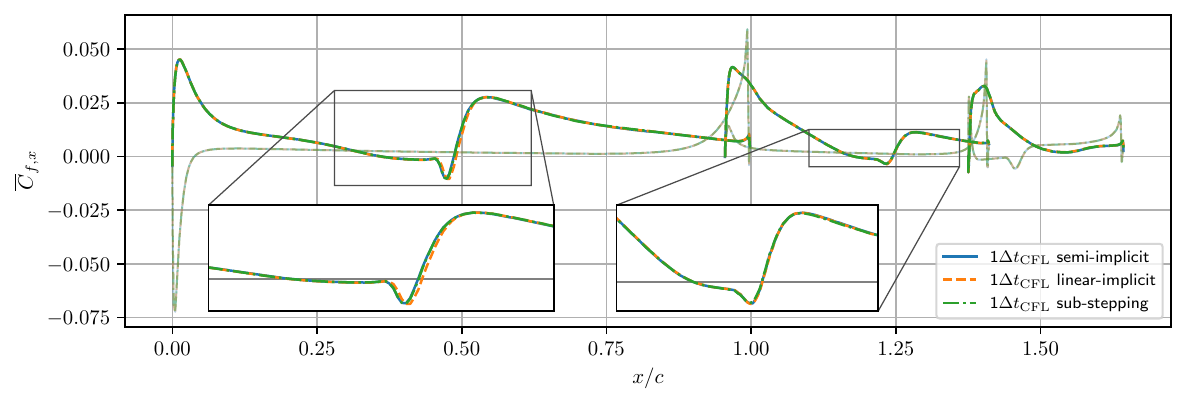}
        \caption{\rev{Streamwise skin-friction coefficient $\overline{C}_{f,x}$. Insets: suction side of the main plane and flap~1.}}
        \label{fig.surface-skinfriction-scheme}
    \end{subfigure}
    \caption{Time-averaged pressure and skin-friction coefficients comparing the influence of the time-stepping scheme. Note that both coefficients are averaged in the spanwise direction. \rev{Line shading and insets in (b) as in figure~\ref{fig.verification-cp-cf}.}}
    \label{fig.surface-scheme}
\end{figure}

Figure~\ref{fig.surface-scheme} compares the time-averaged surface pressure and skin-friction coefficients obtained with the semi-implicit, sub-stepping, and linear-implicit schemes at $\Delta t = \dtcfl$.
Overall, the three schemes produce very similar surface distributions on all wing elements.
In particular, the transition region on the main plane over the range $x/c \in [0.45,0.55]$ is well matched, indicating that the sensitive near-wall flow physics is preserved across the time-stepping schemes.

\begin{figure}[!tb]
    \centering
    \includegraphics[width=\textwidth]{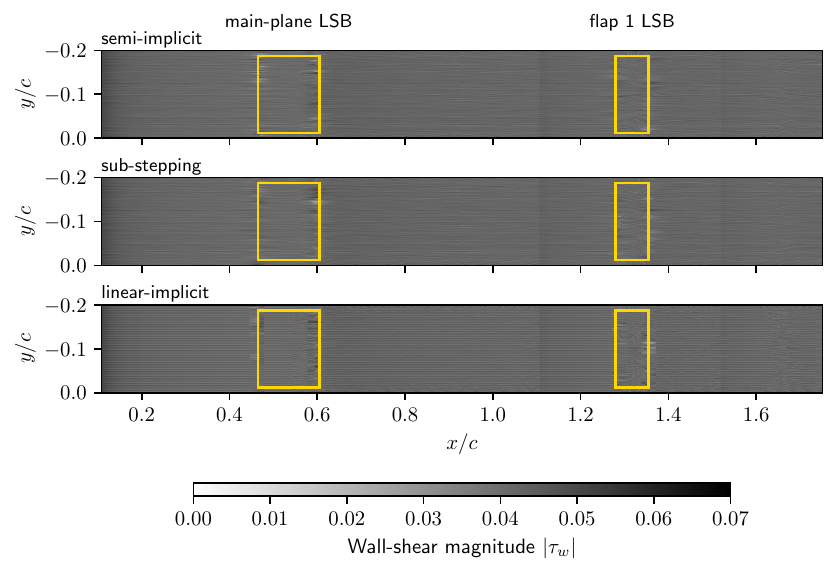}
    \caption{Surface line integral convolution on the suction sides of the wing elements for the \rev{semi-implicit (top), sub-stepping (middle), and linear-implicit (bottom)} schemes at the reference time step size $\dtcfl$\rev{, shaded by the time-averaged wall-shear magnitude. The boxes mark the laminar separation bubbles on the main plane and flap~1}.}
    \label{fig.lic-scheme}
\end{figure}

The skin-friction coefficient in figure~\ref{fig.surface-skinfriction-scheme} shows that the laminar separation bubble on the suction side of the main plane \rev{extends from $x/c \approx 0.36$ to $x/c \approx 0.48$, with the strongest reversed flow at $x/c \approx 0.47$--$0.48$, for all three schemes; the separation and reattachment points differ by at most $0.004c$ between the schemes}.
This is consistent with the pressure recovery observed in figure~\ref{fig.surface-pressure-scheme} at $x/c \approx 0.47$.
Similarly, \rev{on flap~1 all schemes predict a laminar separation bubble between $x/c \approx 1.18$ and $1.24$, followed by} transition\rev{,} as indicated by the rise in both skin friction and pressure.

For flap~2, all three schemes reproduce the same transition pattern, with a laminar separation bubble on the pressure side and bypass transition on the suction side.
The pressure-side \rev{bubble extends from $x/c \approx 1.39$ to $x/c \approx 1.47$}, while the suction side shows a strong increase in skin friction near the leading edge at $x/c \approx 1.38$.

The similarity of the surface flow topology is further illustrated by the line integral convolution (LIC) plots in figure~\ref{fig.lic-scheme}.
\rev{Line integral convolution visualises a vector field by convolving a white-noise texture along its local streamlines \cite{cabral_imaging_1993}, producing a dense, resolution-independent image in which the intensity pattern is smeared along the flow direction and remains uncorrelated across it. Applied here to the time-averaged wall shear-stress field, it depicts the mean skin-friction lines and makes separation and reattachment patterns visible without seeding individual streamlines.}
The suction-side flow structures on the main plane and first flap are reproduced consistently by all three schemes, with little variation in the spanwise direction.
In agreement with \cite{Slaughter2023}, no laminar separation bubble is observed on the suction side of flap~2.
\rev{At $\Delta t = \dtcfl$, the choice of time-stepping scheme therefore has negligible influence on the resolved surface flow physics.}

\subsubsection{Spectral analysis of force and probe signals}
\label{sec.accuracy-scheme-spectra}

\begin{figure}[htbp]
    \centering
    \includegraphics[width=0.7\textwidth]{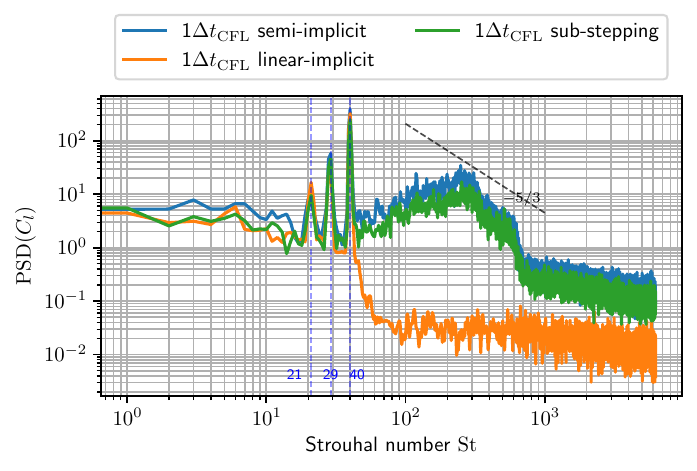}
    \caption{Power spectral density of the lift coefficient summed over all wing elements. Comparison of the influence of the time-stepping scheme with time step size $\dt = \dtcfl$.}
    \label{fig.psd-scheme}
\end{figure}

Figure~\ref{fig.psd-scheme} compares the PSD of the total lift coefficient for the three time-stepping schemes at $\Delta t = \dtcfl$.
All schemes reproduce the dominant low-frequency peaks at $\mathrm{St}=21$, $29$, and $40$ with similar magnitude, indicating that the primary unsteady dynamics are captured consistently.

Differences emerge at higher frequencies.
The semi-implicit and sub-stepping schemes exhibit similar broadband behaviour, including a spectral decay over the range $\mathrm{St}\in[300,600]$.
In contrast, the linear-implicit scheme shows a substantially stronger attenuation for $\mathrm{St}>40$, with the spectrum flattening beyond approximately $\mathrm{St}=80$.
This suggests that the linear-implicit formulation introduces additional numerical damping at higher frequencies even when operated at the reference time step size.

\begin{revblock}
To establish whether this high-frequency attenuation is a spatially localised effect or only a property of the integrated force signal, we complement the force-coefficient PSD with the power spectral density of the resolved turbulent kinetic energy at fixed probes on the main-plane suction side and on the first flap, computed with Welch's method as in section~\ref{sec.verification}.
The main-plane probe is point~\#6 of table~\ref{tab.probe_locations} and the flap~1 probe is point~\#93, both on the \emph{suction} side, see \ref{ap.probes} for details.
\rev{Point~\#6 ($x/c \approx 0.43$) lies within the chordwise extent of the main-plane separation bubble ($0.36 \lesssim x/c \lesssim 0.48$), upstream of reattachment, and point~\#93 ($x/c \approx 1.14$) lies just upstream of the flap~1 bubble ($1.18 \lesssim x/c \lesssim 1.24$).
Both therefore track the growth of the instability ahead of breakdown to turbulence on their element.}

The probe spectra in figure~\ref{fig.psd-probe-scheme} overlap across the whole resolved band; the linear-implicit scheme sits marginally below the other two for $\mathrm{St}\sim150$--$300$, a much weaker signature of the strong high-frequency attenuation seen in the force-coefficient PSD (figure~\ref{fig.psd-scheme}).
Beyond $\mathrm{St}\sim700$ all three schemes flatten into a noise floor, as for the lift-coefficient PSD (section~\ref{sec.verification}); the linear-implicit floor sits roughly two orders of magnitude above the other two\rev{, although the solver tolerances are identical (table~\ref{tab.solverSetup}); we have not investigated its origin}.
The high-frequency deficit of the linear-implicit scheme is therefore only faintly present in the local probe signal at the reference time step size, and much less pronounced than in the lift-coefficient spectrum in figure~\ref{fig.psd-scheme}.

\begin{figure}[htbp]
    \centering
    \begin{subfigure}[b]{0.45\textwidth}
        \centering
        \includegraphics[width=\textwidth]{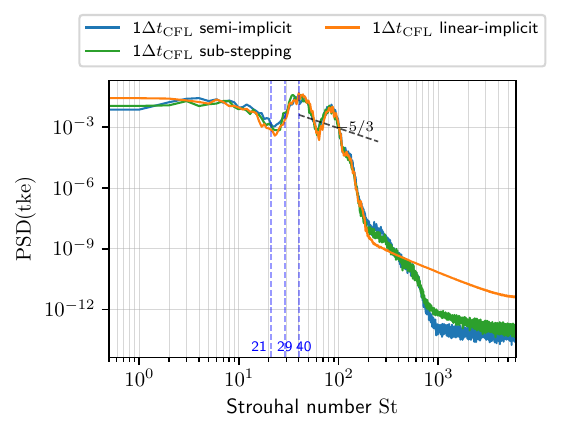}
        \caption{Main plane, suction side.}
        \label{fig.psd-probe-scheme-main}
    \end{subfigure}
    \hfill
    \begin{subfigure}[b]{0.45\textwidth}
        \centering
        \includegraphics[width=\textwidth]{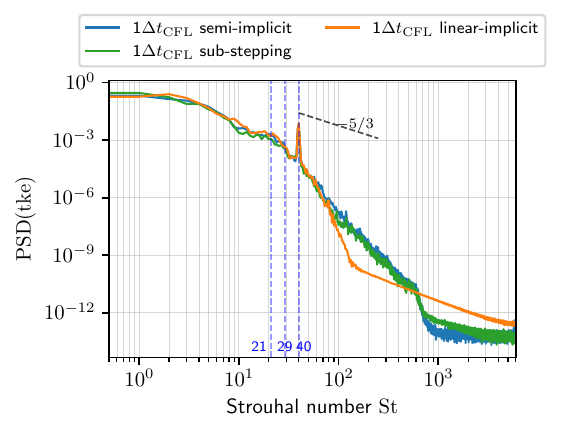}
        \caption{Flap~1, suction side.}
        \label{fig.psd-probe-scheme-flap1}
    \end{subfigure}
    \caption{\rev{Power spectral density of the resolved turbulent kinetic energy at a probe \rev{within the main-plane laminar separation bubble (a, point~\#6) and upstream of the flap~1 bubble (b, point~\#93)}, comparing the semi-implicit, sub-stepping and linear-implicit schemes at the reference time step size $\dtcfl$.}}
    \label{fig.psd-probe-scheme}
\end{figure}
\end{revblock}

\begin{revblock}
\subsubsection{Instantaneous flow and separation bubbles}
\label{sec.instantaneous-scheme}
\label{sec.bubble-organisation-scheme}

\begin{figure}[htbp]
    \centering
    \includegraphics[width=0.92\textwidth]{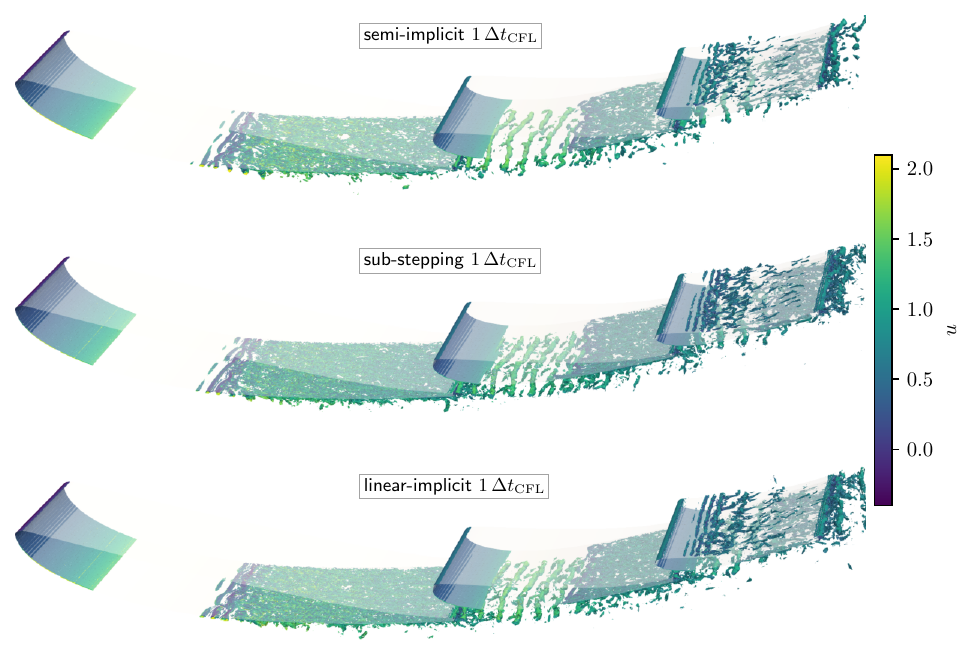}
    \caption{\rev{Instantaneous iso-surface of the $Q$-criterion ($Q=6\times10^{4}$), coloured by streamwise velocity $u$, comparing the semi-implicit, sub-stepping and linear-implicit schemes at the reference time step size $\dtcfl$. The wing elements are shown as semi-transparent surfaces.}}
    \label{fig.qiso-scheme}
\end{figure}

\begin{figure}[htbp]
    \centering
    \includegraphics[width=0.92\textwidth]{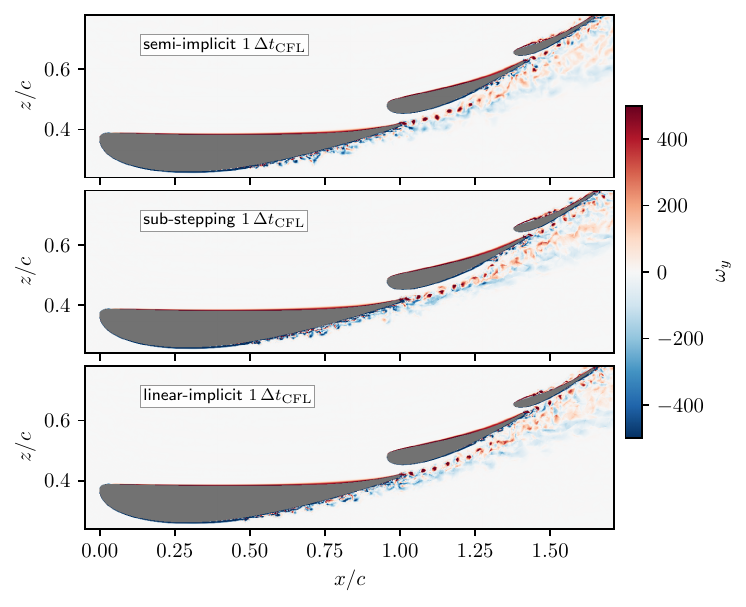}
    \caption{\rev{Instantaneous spanwise vorticity $\omega_y$ on the mid-span plane (\rev{$y=-0.1c$}); the wing cross-sections are shown in grey. Same cases as figure~\ref{fig.qiso-scheme}.}}
    \label{fig.wy-scheme}
\end{figure}

To test whether the choice of scheme is also visible \emph{instantaneously}, which would provide a diagnostic without long statistical sampling, we compare iso-surfaces of the $Q$-criterion, coloured by streamwise velocity, and the spanwise vorticity $\omega_y$ on the mid-span plane.

At $\dt = \dtcfl$ the three schemes produce qualitatively similar instantaneous fields, see figures~\ref{fig.qiso-scheme} and~\ref{fig.wy-scheme}; the snapshots differ in detail, as expected for independent turbulent realisations, but show the same structures.
The suction-side shear layer on each wing element rolls up into spanwise vortices that undergo a secondary three-dimensional instability and break down to fine-scale turbulence at the same chordwise station for all schemes.

\begin{figure}[htbp]
    \centering
    \includegraphics[width=0.92\textwidth]{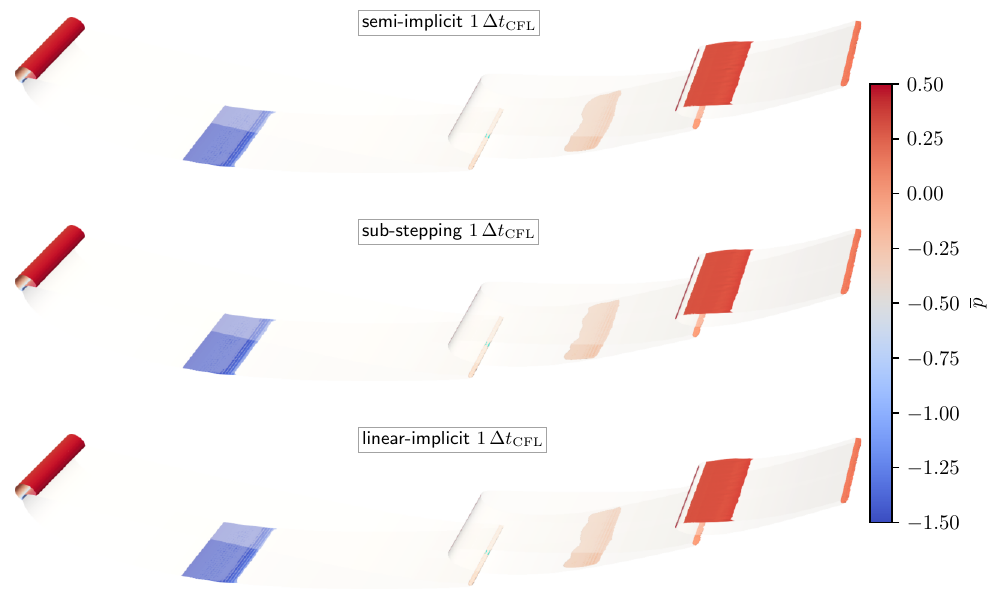}
    \caption{\rev{Three-dimensional organisation of the mean separation bubbles ($\overline{u}=-10^{-4}$ iso-surface, coloured by mean pressure $\overline{p}$, wing elements semi-transparent), comparing the semi-implicit, sub-stepping and linear-implicit schemes at the reference time step size $\dtcfl$.}}
    \label{fig.bubble-organisation-scheme}
\end{figure}

The three-dimensional \emph{mean} flow topology is visualised through an iso-surface of the mean streamwise velocity $\overline{u}$ around all three wing elements, coloured by mean pressure $\overline{p}$.
Because the near-wall value of $\overline{u}$ coincides with the no-slip condition at many mesh nodes, contouring at exactly $\overline{u}=0$ is numerically degenerate (many mesh nodes share the iso-value, which is ill-conditioned for the underlying contouring algorithm); a small negative offset, $\overline{u}=-10^{-4}$, is used instead to reliably select genuinely reversed flow.

Figure~\ref{fig.bubble-organisation-scheme} shows that the three schemes produce the same bubble organisation at $\dtcfl$.
The thin transitional bubble on the main plane (\rev{$0.36 \lesssim x/c \lesssim 0.48$}) and the separation regions on flap~1 and flap~2, consistent with the transition locations from the surface pressure and skin-friction coefficients in section~\ref{sec.accuracy-scheme-surface}, all agree closely in extent, shape and position across schemes.
\end{revblock}

\subsection{Influence of increasing time step size}\label{sec.accuracy-dt}
We next discuss the influence of increasing the time step size $\dt > \dtcfl$ for the implicit time-stepping schemes and compare against the reference time step size $\dtcfl$.
\rev{The linear-implicit results at $20\,\dtcfl$ and $100\,\dtcfl$ use equal-order spaces, see section~\ref{sec.stability}, so their differences from the reference cannot be attributed to the time step size alone.}

\subsubsection{Time-averaged force coefficients}
\label{sec.accuracy-dt-forces}
We return to figure \ref{fig.mean-force-coefficients-total} which shows an increase in the time step size $\dt$ leads to a consistent change in the forces.
\rev{As in section~\ref{sec.accuracy-scheme}, all differences are given relative to the semi-implicit reference at $\dtcfl$.}
The sub-stepping scheme predicts a maximum change of \rev{$2.68\%$} in the mean drag and $0.81\%$ in mean lift force when increasing the time step size from $\dtcfl$ to $20 \dtcfl$.
The force trend is consistent with an upstream shift of the transition location leading to weaker pressure on the suction side and, hence, \rev{a lower downforce magnitude $|\overline{C}_l|$, and a reduction in drag at $20 \dtcfl$}.

The standard deviation shows an increase in magnitude of oscillations around the mean at larger time step sizes.
The relative increase in standard deviation around the mean is from $0.56\%$ at $\dtcfl$ to $3.39\%$ at $20 \dtcfl$ for lift.
\rev{The computations remain bounded and free of grid-scale oscillations at all admissible time step sizes, so the growth in the force fluctuations is not a symptom of marginal numerical stability. It is specific to the sub-stepping scheme: the relative fluctuations of the linear-implicit scheme remain close to the reference (see below), even at $100\dtcfl$, where its resolved flow has become quasi-two-dimensional (section~\ref{sec.instantaneous-dt}). The growth therefore reflects a loss of temporal accuracy of the sub-stepping solution whose origin we have not analysed further.}
Overall, the sub-stepping scheme gives robust predictions of global forces despite large increases in time step size.

In comparison, the linear-implicit scheme shows a gradual increase in the \rev{downforce magnitude $|\overline{C}_l|$} up to $100 \dtcfl$ and a slight reduction in the drag force prediction.
\rev{The maximum difference in the mean is $4.83\%$ for the drag and $1.17\%$ for the lift coefficient, which remain close to the reference prediction.}
Interestingly, the standard deviation \rev{relative to the mean} does not show as strong changes for the linear-implicit scheme with at most $0.54\%$ for lift and $6.10\%$ for drag\rev{, both at $100 \dtcfl$}.
All in all, the mean force coefficients indicate a gradual change with increasing time step size for both time-stepping schemes.

\begin{figure}[htbp]
    \centering
    \begin{subfigure}[b]{\textwidth}
        \centering
        \includegraphics[width=\textwidth]{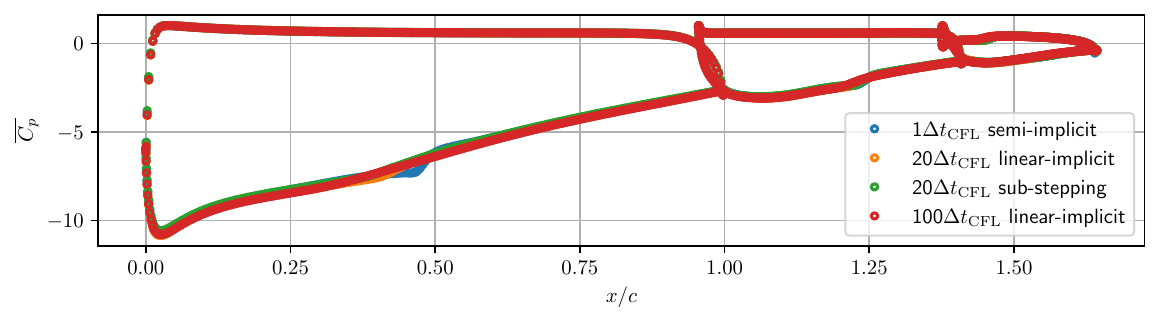}
        \caption{Pressure coefficient $\overline{C}_p$.}
        \label{fig.surface-pressure-dt}
    \end{subfigure}
    \newline
    \begin{subfigure}[b]{\textwidth}
        \centering
        \includegraphics[width=\textwidth]{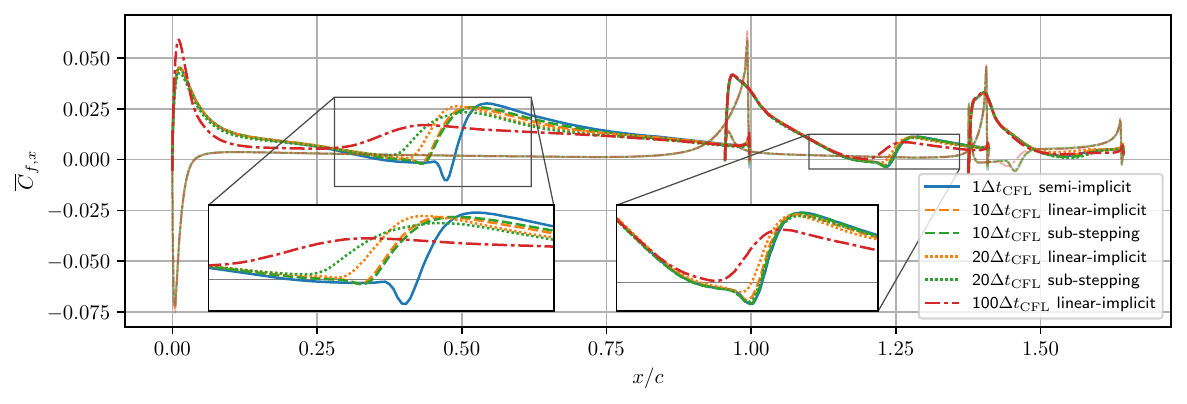}
        \caption{\rev{Streamwise skin-friction coefficient $\overline{C}_{f,x}$. Insets: suction side of the main plane and flap~1.}}
        \label{fig.surface-skinfriction-dt}
    \end{subfigure}
    \caption{Time-averaged pressure and skin-friction coefficients comparing the influence of increasing time step size. \rev{Line shading and insets in (b) as in figure~\ref{fig.verification-cp-cf}. Panel (b) additionally includes $10\dtcfl$ for both implicit schemes; colour indicates the scheme and line style the time step size.} \eqordernote}
    \label{fig.surface-dt}
\end{figure}

\subsubsection{Time-averaged surface coefficients}
\label{sec.accuracy-dt-surface}
Since the changes are gradual from $\dtcfl$ to $20 \dtcfl$, we will \rev{mainly} discuss the increase to $\dt = 20 \dtcfl$ and, separately, the increase to $100 \dtcfl$\rev{; $10\dtcfl$ is included in the skin friction only, to show the loss of the main-plane separation bubble}.
We find that the shift in force coefficients at $20 \dtcfl$ is a consequence of the transition location moving upstream.
Figure \ref{fig.surface-skinfriction-dt} shows the skin-friction coefficient for the sub-stepping and linear-implicit schemes at $\dt = 20 \dtcfl$.
We observe a change in the transition location\rev{, taken as the onset of the skin-friction rise,} upstream from $x/c \approx \rev{0.47}$ to $x/c \approx 0.38$ for the sub-stepping and $x/c \approx \rev{0.40}$ for the linear-implicit scheme.
\rev{The signed skin friction in the inset of figure~\ref{fig.surface-skinfriction-dt} shows how the main-plane laminar separation bubble is lost with increasing time step size. At $\dtcfl$, the suction-side flow reverses ($\overline{C}_{f,x}<0$) over $0.36 \lesssim x/c \lesssim 0.48$, i.e.\ over $0.12c$, with a minimum of $\overline{C}_{f,x} = -0.010$. At $10\dtcfl$, the bubble is still resolved by both implicit schemes, but it is shorter and shallower: it extends over $0.025c$ ($0.413 \le x/c \le 0.437$) for the linear-implicit and $0.033c$ ($0.404 \le x/c \le 0.437$) for the sub-stepping scheme, with a minimum of $\overline{C}_{f,x} \approx -0.002$ for both. At $20\dtcfl$, $\overline{C}_{f,x}$ remains positive upstream of transition on the main-plane suction side for both schemes, so the mean flow no longer separates before transition. This is consistent with the three-dimensional bubble organisation in section~\ref{sec.bubble-organisation-dt}.}
Additionally, the magnitude of the skin-friction peak after transition to turbulence reduces from $\rev{\overline{C}_{f,x} = 0.0276}$ for the semi-implicit scheme by $\rev{15.6} \%$ and $\rev{5.1} \%$ for the sub-stepping and linear-implicit schemes, respectively.
Interestingly, flap~1 and flap~2 do not show a significant shift in the transition location for either time-stepping scheme with $20 \dtcfl$.
\rev{The separation and reattachment points of the suction-side LSB on flap~1 and the pressure-side LSB on flap~2 differ from the reference by at most $0.011c$, at the flap~1 reattachment point of the linear-implicit scheme.}
These observations are consistent with the pressure distribution in figure \ref{fig.surface-pressure-dt} where both time-stepping schemes show an earlier increase in the suction side pressure around $x/c \approx 0.37$ for the sub-stepping and $x/c \approx 0.40$ for the linear-implicit scheme.
\rev{For the sub-stepping scheme, the upstream shift in transition leads to a lower pressure force on the profiles, consistent with the reduced downforce magnitude. For the linear-implicit scheme, the downforce magnitude instead increases slightly despite a similar upstream shift, indicating that the integral lift is not determined by the main-plane transition location alone.}
Similarly, the weaker skin friction at increased time step size coincides with the reduction in the drag coefficient.

The linear-implicit scheme allows an increase to up to $100 \dtcfl$.
While the mean lift and drag coefficients showed only minor changes, the skin friction in figure \ref{fig.surface-skinfriction-dt} highlights a strong difference.
The flow on the main plane suction side has lower skin friction than with $20 \dtcfl$ and shows a shallow increase in skin friction far upstream at $x/c \approx 0.25$.
\rev{Instead of the sharp rise in skin friction at transition observed at $\dtcfl$ and $20 \dtcfl$, the skin friction increases gradually, which suggests a different transition mechanism on the main plane.}
\rev{On flap~1, the suction-side separation bubble disappears ($\overline{C}_{f,x}$ remains positive), and the skin-friction rise moves upstream from $x/c \approx 1.25$ to $1.23$ with a lower peak ($0.0086$ compared to $0.0113$). On flap~2, the pressure-side bubble persists but is weaker and shorter, reattaching at $x/c \approx 1.46$ instead of $1.47$.}
Overall, the time-averaged fields at $100 \dtcfl$ show no large differences in the drag and lift coefficient, but strong changes in the flow structure on all wing elements.

\subsubsection{Spectral analysis of force and probe signals}
\label{sec.accuracy-dt-spectra}
\begin{figure}[htbp]
    \centering
    \includegraphics[width=0.7\textwidth]{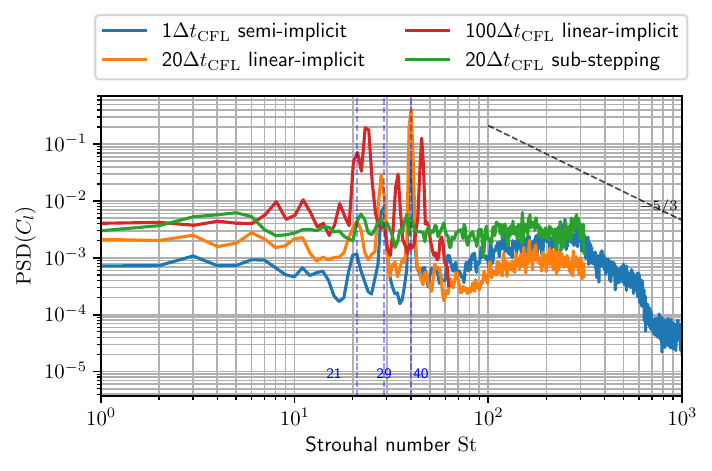}
    \caption{Power spectral density of the total lift coefficient. Comparison of the influence of increasing time step size $\dt > \dtcfl$ for the sub-stepping and linear-implicit scheme. \eqordernote}
    \label{fig.psd-dt}
\end{figure}

We next discuss the PSD of the total lift signal shown in figure~\ref{fig.psd-dt}.
We focus the discussion on the effects for both implicit schemes at $20 \dtcfl$ and, for the linear-implicit scheme, at $100 \dtcfl$.

\rev{The sub-stepping scheme has a broadband spectrum at $20 \dtcfl$ that sits several times \emph{above} the $\dtcfl$ reference level for $\mathrm{St} \lesssim 20$ and again for $\mathrm{St} \gtrsim 60$, rather than being uniformly damped.
Of the reference peaks, only the one at $\mathrm{St} = 21$ remains distinguishable; the band spanning $\mathrm{St} = 29, 40$ instead falls below the reference level.
In contrast, the linear-implicit scheme at $20 \dtcfl$ reproduces a clearly amplified peak near $\mathrm{St} = 40$ and a smaller one near $\mathrm{St} = 21$, though the peak at $\mathrm{St} = 29$ is not distinct against its own elevated broadband content; unlike sub-stepping, its spectrum falls below the reference for $\mathrm{St} > 40$.}
Note that the large increase in time step size to $20 \dtcfl$ leads to a cut-off Strouhal number at \rev{$\mathrm{St} = 625$ using the Nyquist criterion, which lies} within the $-5/3$ decaying spectrum.

The PSD for an increase to $100 \dtcfl$ with the linear-implicit scheme shows strong changes in the frequency domain.
On the one hand, the spectrum captures unsteady characteristics up to \rev{$\mathrm{St} = 125$}, which is well within the resolved spectrum of the lift signal.
On the other hand, it does not identify the original frequency peaks at $\mathrm{St} = 29, 40$, but captures intermediate peaks at $\mathrm{St} = 21, 24, 34, 45$.
The change in the frequency spectrum suggests a change in the instantaneous flow fields for the large increase in time step size.

\begin{revblock}
The same picture is obtained from the temporal spectra of the resolved turbulent kinetic energy at a fixed probe on the first flap, the same one introduced in section~\ref{sec.accuracy-scheme-spectra}, and at a main-plane probe moved further upstream, to point~\#2 of table~\ref{tab.probe_locations} (figure~\ref{fig.psd-probe-dt}).
The main-plane probe is relocated from point~\#6 because increasing the time step size shifts the transition front upstream past it for the sub-stepping and linear-implicit schemes at $20\dtcfl$ (section~\ref{sec.accuracy-dt-surface})\rev{; point~\#2 ($x/c \approx 0.35$) remains upstream of the transition onset at $20\dtcfl$ ($x/c \approx 0.38$--$0.40$), so the comparison still tracks the pre-transition travelling-wave instability rather than the already-turbulent flow behind it. At $100\dtcfl$, where the skin friction rises gradually from $x/c \approx 0.25$, point~\#2 lies within this gradual rise.}
All four cases \rev{(the reference, both implicit schemes at $20\dtcfl$ and the linear-implicit scheme at $100\dtcfl$)} still show a distinguishable peak near $\mathrm{St}\approx40$, though its magnitude relative to the $\dtcfl$ reference varies by more than an order of magnitude between schemes and probes.
Beyond the peak, the spectrum is truncated at each case's Nyquist limit ($\mathrm{St} \approx 625$ at $20\dtcfl$, $\mathrm{St} \approx 125$ at $100\dtcfl$) and the inertial-range energy is progressively lost.
Thus the influence of the time step size, which is only weakly visible in the time-averaged $\overline{C}_l$ and $\overline{C}_d$, is explicit in the probe spectra.

\begin{figure}[htbp]
    \centering
    \begin{subfigure}[b]{0.45\textwidth}
        \centering
        \includegraphics[width=\textwidth]{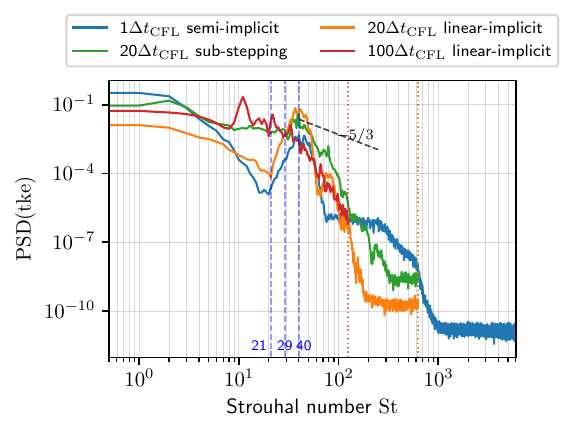}
        \caption{Main plane, suction side.}
        \label{fig.psd-probe-dt-main}
    \end{subfigure}
    \hfill
    \begin{subfigure}[b]{0.45\textwidth}
        \centering
        \includegraphics[width=\textwidth]{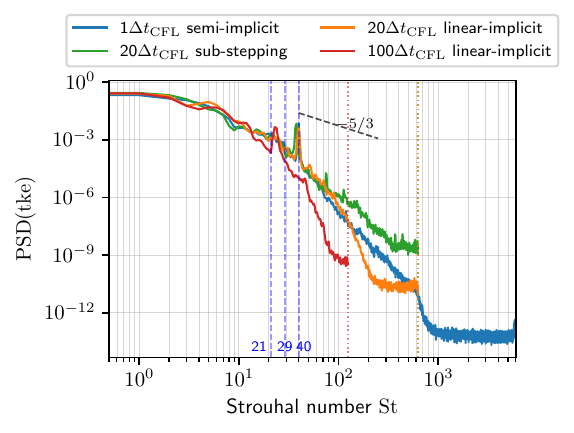}
        \caption{Flap~1, suction side.}
        \label{fig.psd-probe-dt-flap1}
    \end{subfigure}
    \caption{\rev{Power spectral density of the resolved turbulent kinetic energy at the same flap~1 probe as figure~\ref{fig.psd-probe-scheme} (b), and at a main-plane probe moved upstream to point~\#2 of table~\ref{tab.probe_locations} to stay ahead of the shifted transition front \rev{at $20\dtcfl$} (a), comparing the schemes at increasing time step size (semi-implicit at $\dtcfl$ shown as reference). Dotted vertical lines mark the Nyquist Strouhal number of each case.} \eqordernote}
    \label{fig.psd-probe-dt}
\end{figure}
\end{revblock}

\begin{revblock}
\subsubsection{Instantaneous flow and separation bubbles}
\label{sec.instantaneous-dt}
\label{sec.bubble-organisation-dt}

We now apply the instantaneous diagnostic introduced in section~\ref{sec.instantaneous-scheme} ($Q$-criterion iso-surfaces and mid-span spanwise vorticity $\omega_y$) to the effect of increasing the time step size for the two implicit schemes, using instantaneous three-dimensional checkpoint fields\rev{, which are written once per convective time unit}.
At $\dt=\dtcfl$ the three schemes were shown to be qualitatively similar; we examine here how, and at what time step size, this agreement breaks down.

\begin{figure}[htbp]
    \centering
    \includegraphics[width=0.92\textwidth]{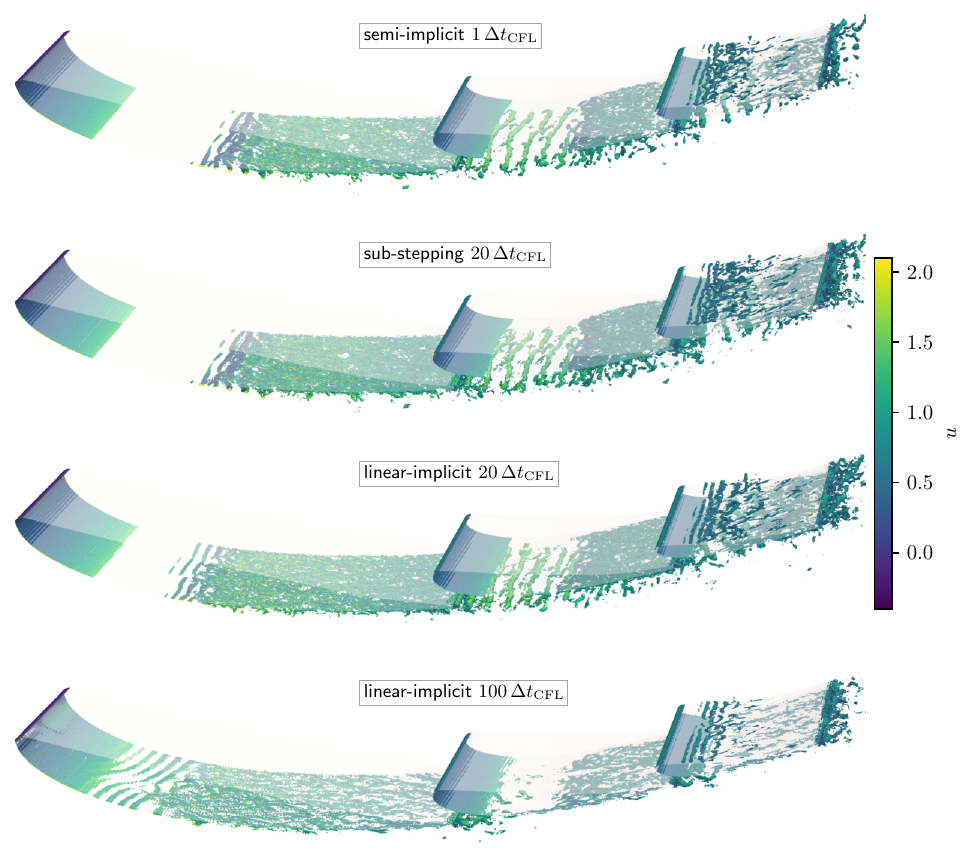}
    \caption{\rev{Instantaneous iso-surface of the $Q$-criterion ($Q=6\times10^{4}$), coloured by streamwise velocity $u$, for the semi-implicit scheme at $\dtcfl$ (reference, cf.\ figure~\ref{fig.qiso-scheme}) and the sub-stepping and linear-implicit schemes at large time step size. The wing elements are shown as semi-transparent surfaces.} \eqordernote}
    \label{fig.qiso-dt}
\end{figure}

\begin{figure}[htbp]
    \centering
    \includegraphics[width=0.92\textwidth]{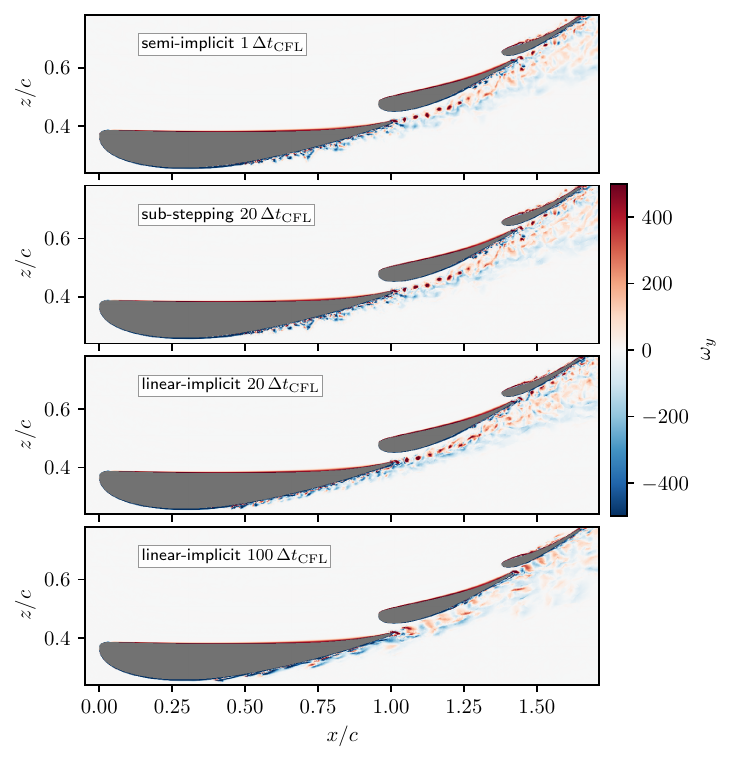}
    \caption{\rev{Instantaneous spanwise vorticity $\omega_y$ on the mid-span plane (\rev{$y=-0.1c$}); the wing cross-sections are shown in grey. Same cases as figure~\ref{fig.qiso-dt}.}}
    \label{fig.wy-dt}
\end{figure}

As the time step size is increased the instantaneous field degrades progressively, and the degradation is apparent in a single snapshot (figures~\ref{fig.qiso-dt} and~\ref{fig.wy-dt}).
At $20 \dtcfl$ the fine-scale structures in both $Q$ and $\omega_y$ are visibly coarser for the sub-stepping and linear-implicit schemes, and the transition front on the main plane has moved upstream, consistent with the upstream shift of the time-averaged skin-friction transition in figure~\ref{fig.surface-skinfriction-dt}.
At $100 \dtcfl$ the linear-implicit field is qualitatively different: the suction-side shear layers on all three elements are dominated by large, spanwise-coherent rollers that persist far downstream, while the three-dimensional secondary instability and the small-scale turbulent cascade are largely suppressed.
In the mid-span $\omega_y$ field the shear layer becomes a single smooth large-amplitude wave, and the onset of small-scale turbulent breakdown is visibly delayed further downstream relative to the reference time step size.
The resolved flow has reverted towards a quasi-two-dimensional state.

The degradation is gradual and the computations remain bounded and free of grid-scale oscillations, so it reflects a loss of temporal accuracy rather than an approaching numerical instability.
It does not by itself explain the larger force fluctuations of the sub-stepping scheme (section~\ref{sec.accuracy-dt-forces}): the linear-implicit solution at $100\dtcfl$ is the most two-dimensional case shown, yet its lift fluctuations remain close to the reference.

\begin{figure}[htbp]
    \centering
    \includegraphics[width=0.92\textwidth]{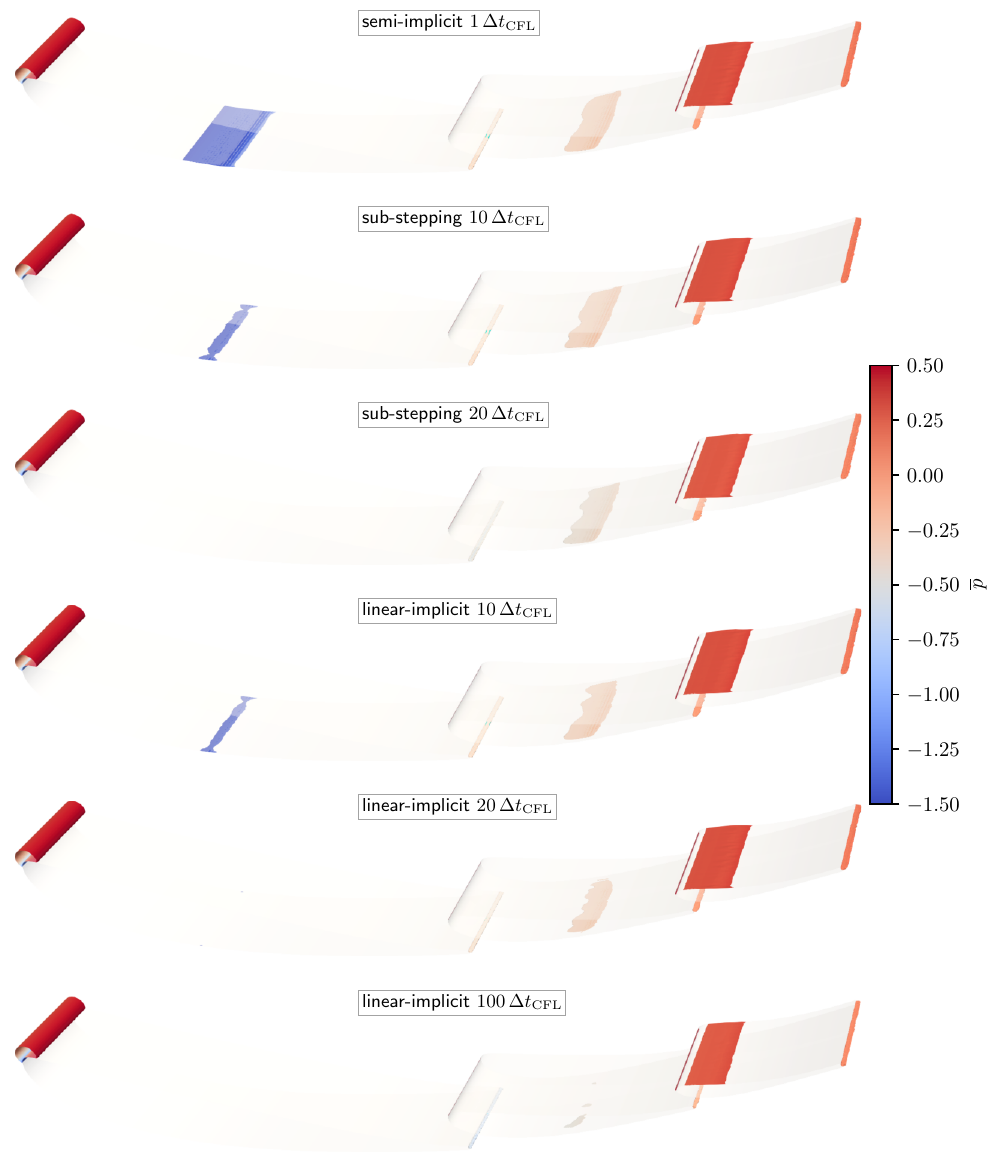}
    \caption{\rev{Three-dimensional organisation of the mean separation bubbles ($\overline{u}=-10^{-4}$ iso-surface, coloured by mean pressure $\overline{p}$, wing elements semi-transparent), for the semi-implicit scheme at $\dtcfl$ (reference) and the sub-stepping and linear-implicit schemes at $10\dtcfl$ and $20\dtcfl$, and additionally the linear-implicit scheme at $100\dtcfl$, the largest stable step tested for either implicit scheme.} \eqordernote}
    \label{fig.bubble-organisation-dt}
\end{figure}

The mean separation bubbles, visualised as in section~\ref{sec.bubble-organisation-scheme}, show that the different mean-flow features respond very differently to increasing time step size (figure~\ref{fig.bubble-organisation-dt}).
The leading-edge stagnation points (section~\ref{sec.meanflow}) are unchanged at every time step size tested for both implicit schemes.

The main, transitional separation bubble on the main plane (\rev{$0.36 \lesssim x/c \lesssim 0.48$} from the skin-friction data of section~\ref{sec.accuracy-scheme-surface}) behaves very differently: it is visibly smaller, but still present, at $10\dtcfl$ for \emph{both} the sub-stepping and linear-implicit schemes, and has disappeared entirely by $20\dtcfl$ for both.
This refines the picture given by the spanwise-averaged skin friction, in which a spanwise-non-uniform bubble can be partially masked by the averaging; the full-span view shows the two implicit schemes losing this bubble at the same rate, consistent with, and a mean-flow counterpart to, the upstream shift of the transition front and the loss of small-scale content documented above for the instantaneous field.
At $100\dtcfl$ (linear-implicit only, the largest stable time step size tested) the main-plane bubble remains absent.

The separation regions on flap~1 and flap~2, identified from the surface coefficients in sections~\ref{sec.accuracy-scheme-surface} and~\ref{sec.accuracy-dt-surface}, retain recognisable surface signatures up to $20\dtcfl$ for both schemes.
At $100\dtcfl$, the flap~1 region also visibly changes character (its pressure signature reverses sign relative to every other case shown)\rev{: the span-averaged skin friction no longer changes sign, but individual spanwise cuts still show short regions of reversed flow at $1.19 \lesssim x/c \lesssim 1.21$, which the iso-surface captures}.
\end{revblock}

\section{Computational performance}
\label{sec.performance}

We assess computational performance through strong scaling, time-to-solution
at larger time steps, and time-to-solution under stated accuracy targets.
The strong-scaling comparison quantifies parallel efficiency and
communication overhead for the three schemes; the latter two comparisons
show the performance gain available before and after imposing accuracy
constraints.

The accuracy analysis in section~\ref{sec.accuracy} showed that \rev{the mean force coefficients of the implicit schemes remain close to the reference up to approximately $20 \dtcfl$, while the main-plane separation bubble is already substantially shortened at $10 \dtcfl$ and absent from the mean flow at $20 \dtcfl$.}
The present section therefore examines whether these larger stable time steps also translate into meaningful computational savings.

Performance measurements are carried out using restart fields taken from a quasi-steady state in order to exclude the highly variable initial transient.
They therefore measure the cost per simulated time interval in that regime,
not the wall-clock time needed to reach it from the initial condition.
\begin{revblock}
The time-to-solution comparisons in sections~\ref{sec.stability-constrained-performance} and~\ref{sec.accuracy-constrained-performance} use the $8$\ctu{} production runs from which the statistics in section~\ref{sec.accuracy} are computed.
The average computational time per time step, $\tdt$, and the average iteration counts are taken over these runs, excluding the first $10\%$ of each run.
The timings include the output of the force, probe and time-averaged field data.
\end{revblock}

All performance measurements were obtained on the ARCHER2 UK national supercomputing service \cite{beckett_archer2_2024}.
Each compute node provides 128 physical cores across two AMD EPYC Rome 7742 processors.
The CPU architecture supports AVX2 vectorisation, which is exploited by \nektarpp together with element grouping for compute-intensive operators \cite{Cantwell2011}.
The code was compiled using GCC 11.3.0 and Cray MPICH 8.1.27.
\rev{All time-to-solution comparisons in sections~\ref{sec.stability-constrained-performance} and~\ref{sec.accuracy-constrained-performance} use 32 nodes (4096 cores). Because the schemes scale differently (section~\ref{sec.strong-scaling}), the relative speed-ups depend on this core count.}

\subsection{Strong scaling}
\label{sec.strong-scaling}

Strong scaling is assessed using the average computational time per time step, $\tdt(N_P)$, as a function of the number of processors $N_P$\rev{, measured in dedicated runs of $100$ time steps at each node count and averaged after the first $10$ steps}.
Since the eIFW case requires at least four ARCHER2 nodes for memory capacity, the configuration with $N_P=512$ processors is used as the baseline.
The speed-up is defined as
\begin{equation}
    S(N_P) = \frac{\tdt(512)}{\tdt(N_P)},
\end{equation}
and the parallel efficiency as
\begin{equation}
    \epsilon_P = \frac{512\,\tdt(512)}{N_P\,\tdt(N_P)}.
\end{equation}

\begin{figure}[htbp]
    \centering
    \includegraphics[width=0.7\textwidth]{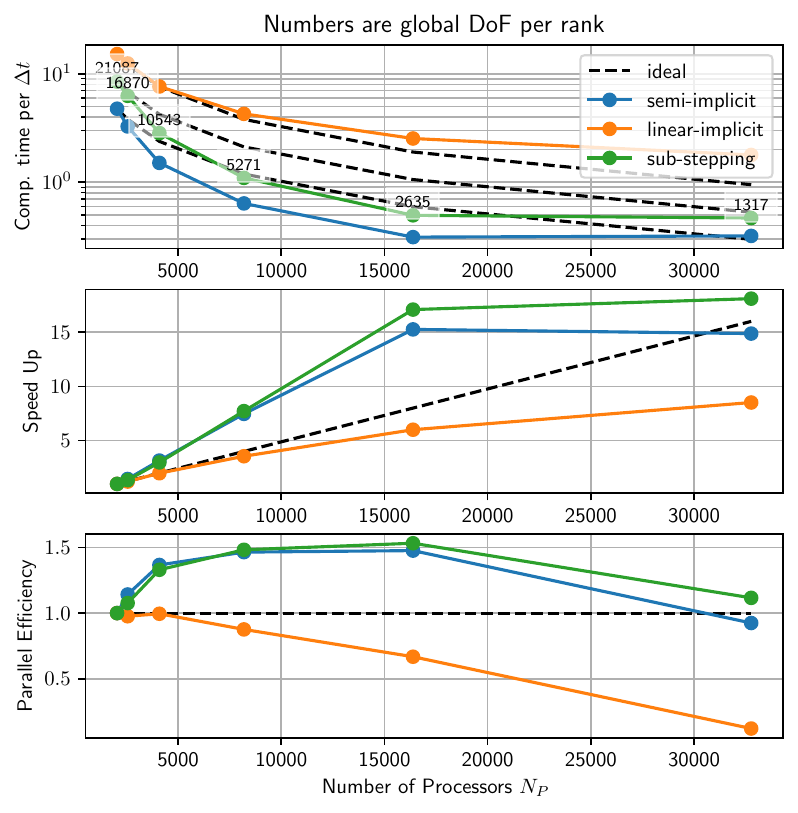}
    \caption{Strong scaling of the semi-implicit, sub-stepping, and linear-implicit schemes in terms of average time per time step, speed-up, and parallel efficiency.}
    \label{fig.scaling-cputime}
\end{figure}

Figure~\ref{fig.scaling-cputime} shows the strong scaling behaviour in terms of time per time step, speed-up, and parallel efficiency.
The semi-implicit and sub-stepping schemes exhibit superlinear scaling over part of the processor range.
This behaviour is attributed to improved cache utilisation and reduced memory pressure as the local problem size per MPI rank decreases.
In particular, the sparse matrix-vector products arising from static condensation become less memory-bound at smaller local degrees of freedom per rank.
Similar behaviour has previously been reported for the compressible solver in \nektarpp \cite{Lindblad2024}.

The linear-implicit scheme shows weaker strong scaling than the semi-implicit and sub-stepping schemes.
\rev{Its time per step is dominated by the GMRES solution of the non-symmetric ADR systems in the velocity step, which accounts for $83$--$96\%$ of the combined pressure and velocity solve time, and whose parallel efficiency relative to $4$ nodes decreases to $0.70$ at $64$ nodes.}
In contrast to the conjugate gradient solver used for the symmetric Poisson and Helmholtz problems, GMRES requires additional global communication during orthogonalisation, particularly in the modified Gram--Schmidt procedure \cite{Saad2003}\rev{, which we expect to contribute to this; the communication cost was not measured separately}.

\subsection{Stability-constrained performance}
\label{sec.stability-constrained-performance}

\begin{figure}[htbp]
    \centering
    \includegraphics[width=0.7\textwidth]{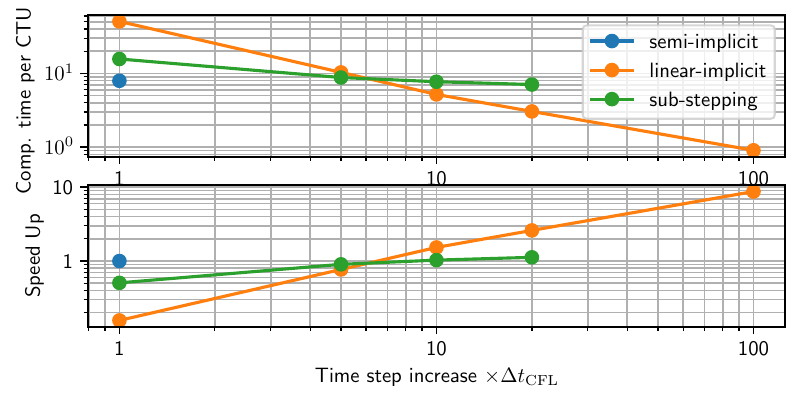}
    \caption{Reduction in time-to-solution relative to the semi-implicit reference as a function of time step size for the sub-stepping and linear-implicit schemes. \eqordernote \rev{Values from the $8$\ctu{} production runs on 4096 cores, see section~\ref{sec.performance}.}}
    \label{fig.implicit-scaling}
\end{figure}

We now evaluate the reduction in time-to-solution that can be achieved by the implicit schemes when the time step size is increased.
For a target physical time interval of one convective time unit (CTU), the time-to-solution is defined as
\begin{equation}
    T_{\mathrm{sol}} = \ndt \tdt,
\end{equation}
where $\ndt = 1\,\mathrm{CTU}/\Delta t$ is the number of time steps required and $\tdt$ is the average computational time per step.

The semi-implicit scheme at $\Delta t = \dtcfl$ is used as the reference configuration.
Figure~\ref{fig.implicit-scaling} reports the resulting speed-up in time-to-solution for the implicit schemes as the time step size is increased.

At the reference time step size, both implicit schemes incur a substantial per-step overhead relative to the semi-implicit formulation.
The sub-stepping scheme is slower by a factor of \rev{$1.97$}, while the linear-implicit scheme is slower by a factor of \rev{$6.37$}.
For the sub-stepping scheme this overhead arises from the additional pseudo-time advection solves.
For the linear-implicit scheme it is caused primarily by reconstruction of the ADR system and the higher cost of solving non-symmetric velocity systems.
Consequently, neither implicit scheme is competitive at $\Delta t = \dtcfl$; their benefit can only be realised if the larger stable time steps reduce the total number of time steps sufficiently.

For the linear-implicit scheme, increasing the time step size leads to a break-even point \rev{between $\Delta t = 5\dtcfl$ and $10\dtcfl$, followed by a reduction in time-to-solution by a factor of $1.53$ at $\Delta t = 10\dtcfl$}.
In this regime, the computational work per time step remains close to that at the reference time step size, so the reduction in the number of time steps translates almost directly into performance gain.

\begin{figure}[htbp]
    \centering
    \includegraphics[width=0.7\textwidth]{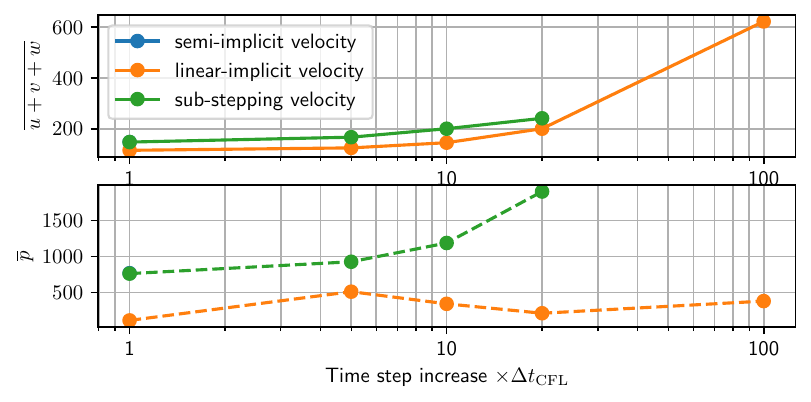}
    \caption{Average iteration counts for the pressure solve and the sum of the three velocity solves as a function of time step size. \eqordernote \rev{Values from the $8$\ctu{} production runs on 4096 cores, see section~\ref{sec.performance}.}}
    \label{fig.implicit-iterations}
\end{figure}

At larger time step sizes, however, the velocity systems become \rev{harder to solve iteratively, as the reaction term $\gamma/\dt$ that makes them diagonally dominant weakens}.
Figure~\ref{fig.implicit-iterations} shows that the total number of iterations over the three velocity solves increases from \rev{$115$} at $\Delta t=\dtcfl$ to \rev{$145$} at $10\dtcfl$, and then rises sharply to \rev{$621$} at $100\dtcfl$.
As a result, the speed-up exhibits diminishing returns and reaches a maximum of \rev{$8.7$} at $\Delta t = 100\dtcfl$.
\rev{The timings at $20\,\dtcfl$ and $100\,\dtcfl$ are for the equal-order discretisation, see section~\ref{sec.stability}, whose pressure system has more unknowns on the same quadrature; the pressure solve cost at these steps is therefore not directly comparable with the Taylor--Hood cases.}

\begin{figure}[htbp]
    \centering
    \includegraphics[width=0.7\textwidth]{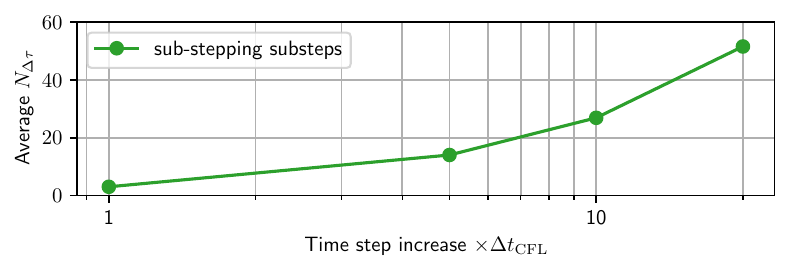}
    \caption{Average number of pseudo-time advection substeps required by the sub-stepping scheme as a function of the outer time step size. \rev{Values from the $8$\ctu{} production runs on 4096 cores, see section~\ref{sec.performance}.}}
    \label{fig.implicit-substeps}
\end{figure}

The sub-stepping scheme exhibits a more modest performance improvement.
A break-even point is reached only at approximately \rev{$\Delta t = 10\dtcfl$, and the time-to-solution is reduced by a factor of $1.12$ at $20\dtcfl$}.
The attainable speed-up is limited by two factors: increasing iteration counts in the pressure and velocity solves, and the growing number of pseudo-time advection substeps required to maintain stability.

As shown in figure~\ref{fig.implicit-iterations}, the pressure iterations increase substantially with time step size and reach an average of \rev{$1900$} at $\Delta t = 20\dtcfl$\rev{. Since the Poisson operator does not depend on $\dt$, this growth presumably reflects the larger change of the pressure between time steps, which makes the previous solution a poorer initial guess for the iterative solver}.
In addition, figure~\ref{fig.implicit-substeps} shows that the average number of pseudo-time substeps grows almost linearly with the outer time step size, from $3$ at $\Delta t=\dtcfl$ to \rev{$26.9$} at $10\dtcfl$ and \rev{$51.6$} at $20\dtcfl$.
This increase follows directly from the growth of the CFL number shown in figure~\ref{fig.cfl-scheme-dt}.
Together, these effects strongly limit the achievable reduction in time-to-solution.

The timing results show that both implicit schemes can reduce the time-to-solution, but through different trade-offs.
The linear-implicit scheme offers the larger potential speed-up, provided that the increase in time step size is sufficient to offset its high per-step overhead.
The sub-stepping scheme has a lower per-step cost but its performance is strongly constrained by the growth in pseudo-time substeps and pressure iterations.
These results show that stability improvements alone are not sufficient to guarantee performance gains; the practical benefit of implicit time-stepping is determined by the combined behaviour of solver cost, iteration growth, and admissible time step size. We next impose accuracy targets on this comparison.

\begin{revblock}
\subsection{Accuracy-constrained performance}
\label{sec.accuracy-constrained-performance}

The speed-ups above account for computational cost but not the changes in flow
physics at larger time steps. For an industrial design task in which the mean
aerodynamic loads are the quantities of practical interest, we constrain the
errors in mean lift and drag and report changes in the surface fields and
separation-bubble footprints separately. We use the semi-implicit solution at
$\dtcfl$ as the reference. For scheme $s$ and time step $\Delta t$, the
relative errors in the mean force coefficients are
\begin{equation}
 e_q(s,\Delta t)=
 \frac{|q(s,\Delta t)-q_{\mathrm{ref}}|}{|q_{\mathrm{ref}}|},
 \qquad q\in\{\overline C_l,\overline C_d\},
\end{equation}
while those in the mean surface distributions are
\begin{equation}
 e_q(s,\Delta t)=
 \frac{\|q(s,\Delta t)-q_{\mathrm{ref}}\|_{L^2(\Gamma)}}
 {\|q_{\mathrm{ref}}\|_{L^2(\Gamma)}},
 \qquad q\in\{\overline C_{f,x},\overline C_p\},
\end{equation}
where $\Gamma$ comprises the three-dimensional wing surfaces. Because a
global norm can miss a local bubble, we also compare the suction-side LSBs
on the main plane and flap~1 and the pressure-side LSB on flap~2. For each
element $i$, the two zero crossings of the negative interval in the
spanwise-averaged $\overline C_{f,x}$ give separation and reattachment
locations $\xi_{s,i}$ and $\xi_{r,i}$, normalised by the main-plane chord
$c=0.25\,\mathrm{m}$ and its leading-edge origin. Let
$\ell_i=\xi_{r,i}-\xi_{s,i}$ and
$\boldsymbol a_i=(\xi_{s,i},\xi_{r,i},\ell_i)$. We define
\begin{equation}
e_{\rm LSB,i}(s,\Delta t)=
\begin{cases}
\displaystyle\frac{\|\boldsymbol a_i-\boldsymbol a_{i,\rm ref}\|_\infty}
 {\ell_{i,\rm ref}}, & \text{both crossings exist},\\[4pt]
1, & \text{otherwise},
\end{cases}
\end{equation}
and $e_{\rm LSB}=\max_i e_{\rm LSB,i}$. The reference length scales errors
in both location and size; a missing crossing pair receives 100\% error.
This is a span-averaged footprint, not a three-dimensional bubble volume.

We define a force-based accuracy metric without allowing agreement in one
force coefficient to compensate for error in the other:
\begin{equation}
 A(s,\Delta t;\tau_l,\tau_d)=\max\!\left\{
 \frac{e_{\overline C_l}(s,\Delta t)}{\tau_l},
 \frac{e_{\overline C_d}(s,\Delta t)}{\tau_d}\right\}.
\end{equation}
Our principal design-oriented comparison uses $\tau_l=0.01$ and
$\tau_d=0.05$, placing a tighter constraint on lift, the principal
aerodynamic load. These are application targets relative to the reference,
not statistical uncertainty bounds; their effect is examined below. A
configuration is accepted for this
force-based comparison when $A(s,\Delta t;\tau_l,\tau_d)\leq1$. The errors in
$\overline C_{f,x}$, $\overline C_p$, and the LSB footprint do not enter this
decision; they reveal changes in the resolved flow that force agreement alone
can conceal.

Table~\ref{tab.accuracy-constrained-performance} evaluates this rule for every
tested configuration while retaining all five relative errors.
Figure~\ref{fig.accuracy-vs-time} shows the same cases in the
$(T_{\mathrm{sol}},A)$ plane and shades the force-accepted region $A\leq1$. At
$10\dtcfl$, the linear-implicit lift and drag errors are 0.51\% and 3.93\%,
respectively, and both satisfy the chosen force targets. Its mean pressure
error is 1.77\%, but its signed skin-friction error is 20.72\% and its
largest LSB-footprint error is 77.93\%. The main-plane bubble is therefore
substantially shortened even though the integrated loads remain close to the
reference. At $20\dtcfl$, the sub-stepping scheme still meets both force
targets, although the selected main-plane reversal is absent; the
linear-implicit scheme exceeds the lift target at $20\dtcfl$ and
$100\dtcfl$. Force-based acceptance should therefore not be interpreted as
agreement in transition-sensitive flow structure.

\begin{table}[H]
    \centering
    \caption{Accuracy and time-to-solution for all tested configurations,
    ordered with the semi-implicit reference first. Errors are relative to
    that reference; $T_{\mathrm{sol}}$ is the wall-clock time per CTU \rev{on 4096 cores}.
    Acceptance requires lift error $\leq1\%$ and drag error $\leq5\%$;
    surface and LSB-footprint errors are diagnostic.
    \rev{$^\dagger$Equal-order spaces ($P_p = P_v = 4$, section~\ref{sec.stability}); errors include the change in pressure polynomial order.}}
    \label{tab.accuracy-constrained-performance}
    \footnotesize
    \setlength{\tabcolsep}{3.5pt}
    \begin{tabular}{lrrrrrrrrc}
        \toprule
        & $\Delta t/\dtcfl$ & \multicolumn{5}{c}{Relative error (\%)}
        & & $T_{\mathrm{sol}}$ & \\
        Scheme & & $\overline C_l$ & $\overline C_d$ & $\overline C_{f,x}$ & $\overline C_p$
        & $e_{\rm LSB}$ & $A$ & (h/CTU) & Accepted \\
        \midrule
        Semi-implicit   & 1   & 0.00 & 0.00 & 0.00 & 0.00 & 0.00 & 0.00 & \rev{7.94} & Yes \\
        \midrule
        Sub-stepping    & 1   & 0.05 & 0.21 & 2.62 & 0.09 & 0.79 & 0.05 & \rev{15.67} & Yes \\
                        & 5   & 0.66 & 0.83 & 14.80 & 1.53 & 40.68 & 0.66 & \rev{8.79} & Yes \\
                        & 10  & 0.81 & 0.39 & 18.79 & 1.83 & 68.48 & 0.81 & \rev{7.71} & Yes \\
                        & 20  & 0.63 & 2.68 & 27.95 & 2.01 & 100.00 & 0.63 & \rev{7.07} & Yes \\
        \midrule
        Linear-implicit & 1   & 0.31 & 1.19 & 3.84 & 0.50 & 4.75 & 0.31 & \rev{50.56} & Yes \\
                        & 5   & 0.36 & 2.44 & 14.39 & 1.32 & 41.74 & 0.49 & \rev{10.32} & Yes \\
                        & 10  & 0.51 & 3.93 & 20.72 & 1.77 & 77.93 & 0.79 & \rev{5.19} & Yes \\
                        & 20$^\dagger$  & 1.17 & 4.69 & 27.05 & 2.43 & 100.00 & 1.17 & \rev{3.05} & No  \\
                        & 100$^\dagger$ & 1.09 & 4.83 & 36.70 & 2.61 & 100.00 & 1.09 & \rev{0.91} & No  \\
        \bottomrule
    \end{tabular}
\end{table}

\begin{figure}[htbp]
    \centering
    \includegraphics[width=\textwidth]{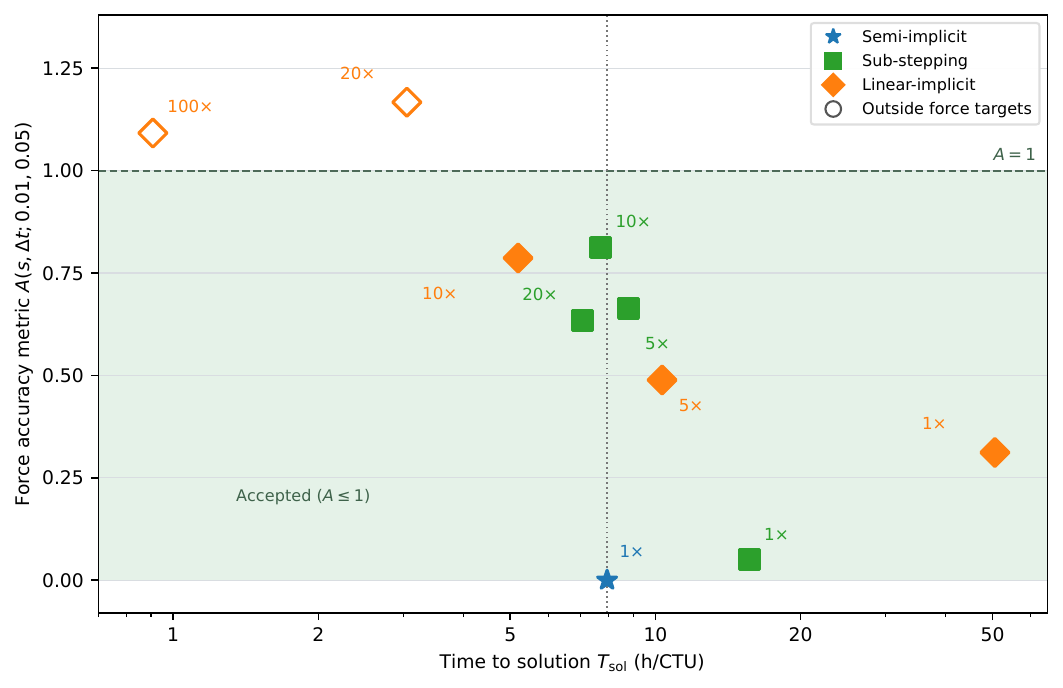}
    \caption{Force-based accuracy metric $A$ versus time-to-solution for all
    tested configurations. The shaded region satisfies the 1\% mean-lift and
    5\% mean-drag targets; filled symbols are accepted and open symbols
    exceed at least one target. Labels give $\Delta t/\dtcfl$ and the dotted
    vertical line marks the semi-implicit reference time. Surface and
    LSB-footprint errors remain in Table~\ref{tab.accuracy-constrained-performance}
    as diagnostics. \eqordernote}
    \label{fig.accuracy-vs-time}
\end{figure}

For each scheme $s$, the accuracy-constrained speed-up is therefore
\begin{equation}
 S_{\mathrm{acc}}^{(s)}=
 \frac{T_{\mathrm{sol,ref}}}
 {\displaystyle\min_{\Delta t:\,A(s,\Delta t;\tau_l,\tau_d)\leq1}
 T_{\mathrm{sol}}^{(s)}(\Delta t)}.
\end{equation}
Within the tested time steps, the least expensive accepted sub-stepping case
is $20\dtcfl$ at \rev{7.07} h/CTU, giving $S_{\mathrm{acc}}=\rev{1.12}$ relative to the
\rev{7.94} h/CTU reference. The least expensive accepted linear-implicit case is
$10\dtcfl$ at \rev{5.19} h/CTU, giving $S_{\mathrm{acc}}=\rev{1.53}$\rev{, a reduction in
time-to-solution by about one third} for the stated mean-force targets.
For comparison, choosing the fastest stable tested case without an accuracy
requirement gives stability-constrained speed-ups of \rev{1.12} for sub-stepping
at $20\dtcfl$ and \rev{8.74}
for linear-implicit at $100\dtcfl$ \rev{(equal-order)}. The force targets leave the
sub-stepping choice unchanged but reduce the admissible linear-implicit
speed-up from \rev{8.74 to 1.53}.
The attainable speed-up depends strongly on those targets. If the drag target
is tightened from 5\% to 3\% while the 1\% lift target is retained,
linear-implicit at $10\dtcfl$ fails because its drag error is 3.93\%; its
fastest accepted case is then $5\dtcfl$ at \rev{10.32} h/CTU
($S_{\mathrm{acc}}=\rev{0.77}$). Sub-stepping at $20\dtcfl$ still passes and gives
$S_{\mathrm{acc}}=\rev{1.12}$. Conversely, relaxing the lift target to 2\% while
retaining 5\% for drag admits linear-implicit at $100\dtcfl$ and its
\rev{$8.74$} speed-up, despite the loss of the main-plane bubble and substantial
skin-friction error. These comparisons are confined to the time steps tested;
they do not imply an interpolated optimum between them.
For gathering quasi-steady mean-force statistics, $10\dtcfl$ with the
linear-implicit scheme is the fastest tested choice satisfying the stated
lift and drag targets. The same choice is insufficient if the requested
statistics depend on the main-plane bubble or transition-sensitive surface
fields; those observables require a separate accuracy constraint.
\end{revblock}

\section{Conclusions}
\label{sec.conclusions}

We have investigated two implicit variants of the velocity correction scheme for incompressible Navier--Stokes simulations on complex geometry, using the extruded Imperial Front Wing benchmark as a representative wall-resolved LES test case.
The study compared a semi-implicit reference formulation with sub-stepping and linear-implicit schemes in terms of stability limits, temporal accuracy, and computational performance.

Both implicit schemes substantially relax the stability restriction of the semi-implicit method.
\rev{Stable simulations were obtained up to $20\dtcfl$ for the sub-stepping scheme and up to $10\dtcfl$ for the linear-implicit scheme with Taylor--Hood spaces, while the linear-implicit scheme remained stable up to $100\dtcfl$ when equal-order approximation spaces were used for velocity and pressure; its results at $20\dtcfl$ and $100\dtcfl$ use this equal-order discretisation.}
At the reference time step size, both implicit schemes produced essentially the same resolved flow physics as the semi-implicit formulation.
As the time step size was increased, the first significant temporal errors appeared in transition-sensitive surface quantities, particularly on the main plane, where the laminar--turbulent transition shifted upstream.
In contrast, the integral force coefficients remained comparatively robust up to approximately $20\dtcfl$.
At $100\dtcfl$, the linear-implicit solution showed substantial changes in both the time-averaged flow field and the dominant spectral content, indicating that the transition dynamics were no longer adequately resolved.
\rev{These errors are also visible directly in the instantaneous flow: as the time step size is increased the resolved vortical structures coarsen and, at $100\dtcfl$, collapse to spanwise-coherent rollers with the small-scale turbulent breakdown suppressed. At $20\dtcfl$ and above, a single instantaneous snapshot therefore reveals that the admissible time step size has been exceeded, without recourse to long statistical sampling. The mean flow shows a corresponding signature: the main transitional separation bubble weakens progressively with time step size, is substantially shortened at $10\dtcfl$ and is absent from the mean flow at $20\dtcfl$. The shortening at $10\dtcfl$ was identified from the time-averaged skin friction; instantaneous fields at this time step size were not examined.}

The computational performance study showed that stability improvements alone do not guarantee lower time-to-solution.
The sub-stepping scheme incurred a moderate per-step overhead, but its speed-up was limited by the near-linear growth in pseudo-time substeps and increasing pressure iterations.
The linear-implicit scheme had a substantially larger per-step overhead due to reconstruction of the ADR systems and the use of GMRES, but it also provided the largest overall reduction in time-to-solution, reaching a speed-up of \rev{$8.7$} at $100\dtcfl$ \rev{with equal-order spaces}.
\rev{This is the maximum among stable tested cases; applying the 1\% mean-lift and 5\% mean-drag targets reduces the achievable linear-implicit speed-up to 1.53 at $10\dtcfl$. For sub-stepping, $20\dtcfl$ remains the fastest tested case under both stability and these force targets, with a speed-up of 1.12. Tightening the drag target to 3\% removes the linear-implicit time-to-solution advantage: its fastest accepted case gives a speed-up of 0.77.}
\rev{The stability-constrained result suggests using large linear-implicit steps to advance through the initial transient before sampling, while the accuracy-constrained result supports $10\dtcfl$ for gathering quasi-steady mean-force statistics under the stated targets.}
\rev{The main-plane separation bubble is already substantially shortened at $10\dtcfl$ (section~\ref{sec.bubble-organisation-dt}) and absent at $20\dtcfl$. Force-based acceptance therefore does not imply preservation of transition-sensitive flow structure.}

Overall, the study demonstrates that implicit velocity correction schemes can provide meaningful reductions in time-to-solution for complex scale-resolving simulations, but only within accuracy limits set by the underlying flow physics.
The most useful operating regime is therefore not defined solely by numerical stability, but by the balance between admissible time step size, acceptable temporal error, and solver cost.

\rev{A natural extension is an adaptive time-stepping strategy that minimises the wall-clock time for the complete simulation: use large stable steps while advancing through the initial transient, reduce the step smoothly as the flow approaches a quasi-steady state, and enforce the accuracy target appropriate to the statistics being sampled. The solution would need to re-equilibrate after the step reduction before averaging begins. Measuring the resulting total time-to-solution, including this adjustment period and the statistical sampling window, would test whether the expected transient saving outweighs the cost of meeting the later accuracy target.}
In addition, the linear-implicit scheme would likely benefit from a matrix-free formulation to reduce the overhead associated with reconstruction of the velocity operators.
Further, a preconditioner for the ADR matrix which targets large time step sizes would further improve the speed-up with time step size.

These findings provide quantitative guidance for selecting implicit time integration strategies in large-scale incompressible flow solvers targeting complex wall-resolved LES configurations.

\section*{Funding Sources}

\begin{minipage}[h]{20mm}
\includegraphics[width=18mm]{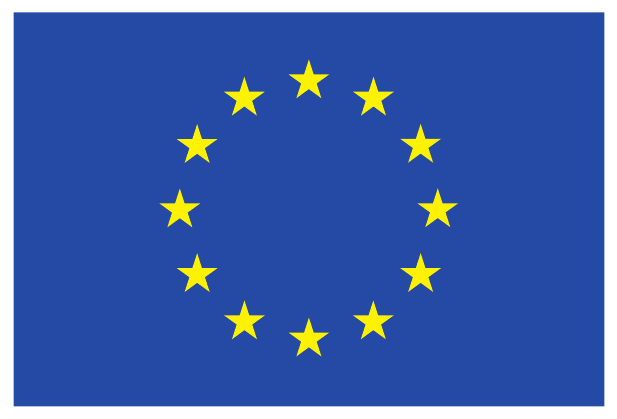}
\end{minipage}\hfill
\begin{minipage}[h]{100mm}
This project received funding from the European Union’s Horizon 2020 research and innovation programme under the Marie Skłodowska-Curie grant agreement No 955923.
\end{minipage} \hfill

\section*{CRediT authorship contribution statement}
\textbf{Henrik Wüstenberg}: Conceptualization, Methodology, Software, Validation, Formal analysis, Investigation, Data Curation, Writing - Original Draft, Writing - Review \& Editing, Visualization, Project administration.
\textbf{Alexandra Liosi}:  Conceptualization, Methodology, Validation, Formal analysis, Investigation, Writing - Review \& Editing, Visualization.
\textbf{Spencer J. Sherwin}: Conceptualization, Software, Writing - Review \& Editing, Supervision, Funding acquisition.
\textbf{Joaquim Peiró}: Conceptualization, Writing - Review \& Editing, Supervision, Project administration, Funding acquisition.
\textbf{David Moxey}: Conceptualization, Software, Writing - Review \& Editing, Supervision, Funding acquisition, Resources.

\section*{Competing Interests}
\rev{The authors declare that they have no known competing financial interests or
personal relationships that could have appeared to influence the work reported
in this paper.}

\section*{Data availability}
The source code for \nektarpp is available at \url{https://www.nektar.info/getting-started/repository/}. Note we produce all results in this work with version 5.8.0. The computational mesh and output data \rev{are available} on the Imperial College London Helix repository under DOI \rev{\url{https://doi.org/10.82186/tpapv-twz85}}.

\section*{Acknowledgement}
Computational resources were supported in part through the Imperial College Research Computing Service (\url{https://doi.org/10.14469/hpc/2232}) and the ARCHER2 UK National Supercomputing Service (\url{https://www.archer2.ac.uk}). David Moxey acknowledges support from the Royal Academy of Engineering under their Research Chair scheme.

\section*{Declaration of generative AI and AI-assisted technologies in the manuscript preparation process}
\rev{During the preparation of this work the authors used ChatGPT, Claude, and Codex in order to enhance the readability of the work. After using these tools, the authors reviewed and edited the content as needed and take full responsibility for the content of the published article.}

\appendix

\section{Iterative tolerance for force spectra resolution}
\label{ap.tolerance}

The iterative tolerance impacts the resolution of small scales within the spectra of force signals.
We identify this in the verification in figure \ref{fig.verification-psd} where we observe a plateau in the high-frequency spectrum at around $\mathrm{St}=700$ while the reference simulation shows a decaying spectrum.
The plateau is a consequence of the iterative tolerance $\epsilon$ of $10^{-4}$ for pressure, see table \ref{tab.solverSetup} for details on iterative solvers.

\begin{figure}[htbp]
    \centering
    \begin{subfigure}[b]{0.45\textwidth}
        \centering
        \includegraphics[width=\textwidth]{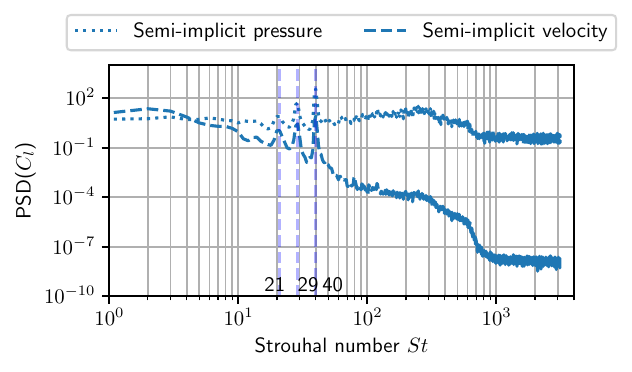}
        \caption{PSD of pressure and viscous lift.}
        \label{fig.tolerance-components}
    \end{subfigure}
    \begin{subfigure}[b]{0.45\textwidth}
        \centering
        \includegraphics[width=\textwidth]{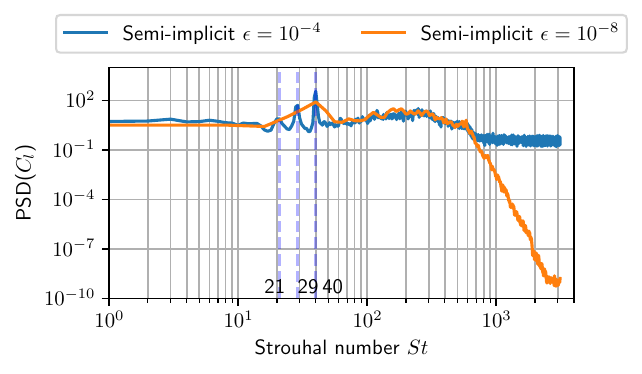}
        \caption{PSD of total lift.}
        \label{fig.tolerance-total}
    \end{subfigure}
    \caption{Power spectral density of the lift force components for the semi-implicit prediction. (a) comparison of pressure- and velocity-driven component (b) variation in iterative pressure tolerance $\epsilon$.}
    \label{fig.tolerance}
\end{figure}

We decompose the total lift force into the pressure and viscous components in figure \ref{fig.tolerance-components} below.
We observe that the velocity spectrum is well-resolved, however, the pressure spectrum shows a plateau at $\mathrm{St} = 700$ and PSD of order $\mathcal{O}(10^0)$ which leads to the plateau in the total coefficient as discussed in the verification section \ref{sec.verification}.
Consequently, we compare the verification case against the same reference simulation however with a lower pressure tolerance of $\epsilon = 10^{-8}$.
We observe that the decay of the high frequency is recovered using a lower pressure tolerance.
The decrease of four orders of magnitude leads to an order $\mathcal{O}(10^{-8})$ decrease in the PSD.

Notably, sufficient iterative tolerances are important for gathering accurate statistics, however, less relevant for the transition.
This encourages mixed precision approaches for potential speed-up of simulations within the initial transient.

\begin{revblock}
To quantify the actual time-to-solution penalty of tightening the pressure tolerance, we repeat the $\epsilon=10^{-8}$ verification run on the same allocation as the $\epsilon=10^{-4}$ semi-implicit case, all other solver settings unchanged, see table \ref{tab.solverSetup}.
Tightening the tolerance increases the mean number of pressure CG iterations per time step from 766 to 3170 and the measured wall-clock cost per time step increases by $127\%$.
Extrapolated to the length of a production run as reported in this paper, this corresponds to an additional $\sim\!3.3\times10^5$ core-hours per case.
For a campaign spanning three schemes and multiple time-step sizes such as the present one, this cost is prohibitive if applied uniformly.

The value $\epsilon=10^{-4}$ used throughout this paper was chosen because it is sufficient for the quantities of interest: the mean wall and force coefficients $\overline{C}_p$, $\overline{C}_f$, $\overline{C}_l$, $\overline{C}_d$.
The dominant transition and load-fluctuation frequencies of the force spectra, $\mathrm{St}\lesssim700$, are unaffected by the solver tolerance, as shown by the \rev{agreement with the reference data in figure~\ref{fig.verification-psd}, the overlap of the three schemes in figure~\ref{fig.psd-scheme} and the tolerance comparison in} figure~\ref{fig.tolerance}.
Only the high-frequency tail of the instantaneous force signal, well above the wing's dominant unsteady frequencies, is contaminated by the algebraic residual noise of the pressure solve at this tolerance.
We therefore recommend $\epsilon=10^{-4}$ as an adequate default for wall-resolved LES campaigns targeting mean loads, wall quantities, and low-to-mid-frequency unsteadiness, and recommend tightening the tolerance only when the instantaneous force or pressure signal itself is the quantity of interest, for instance in aeroacoustic applications.
In that case, the mixed-precision strategy noted above, whereby the tight tolerance is only enforced once the initial transient has passed and statistics are being accumulated, recovers most of the accuracy.

Increasing the level of SVV dissipation is not an effective substitute for tightening the pressure tolerance.
SVV regularises the spatial discretisation by damping energy near the truncation wavenumber of the spectral element basis, targeting aliasing-driven energy pile-up at the resolved-scale cutoff, see section~\ref{sec.svv}.
The high-$\mathrm{St}$ plateau discussed here, in contrast, is an algebraic error floor set by the residual of the iterative pressure solve, independent of spatial resolution.
Increasing SVV dissipation would not act on this error source; it would instead further damp resolved turbulent scales and degrade the very statistics -- transition location, separation, mean loads -- that the tolerance study above shows are already unaffected by the plateau.

A related risk is that the pressure-tolerance error discussed above and the temporal truncation error of a large time step, see section~\ref{sec.accuracy-dt}, could in principle compensate one another in a time-averaged quantity, yielding an apparently unaffected mean statistic that in fact masks two opposing biases.
The two errors remain separable in practice because they act on different parts of the spectrum -- \rev{for the semi-implicit reference,} the tolerance floor only above $\mathrm{St}\approx700$, outside the resolved band -- so the instantaneous diagnostics of sections~\ref{sec.accuracy-dt-spectra} and~\ref{sec.instantaneous-dt} remain a reliable route for detecting an otherwise-hidden, time-step-driven unphysical state even where mean loads appear unaffected.
\end{revblock}

\section{Influence of the Reynolds number}
\label{ap.reynolds}

This appendix examines the influence of the Reynolds number on the benchmark configuration in order to assess whether the discrepancy with the reference setup in \cite{Slaughter2023} can explain differences observed in the verification study.
We compare two otherwise identical semi-implicit simulations with Reynolds numbers \rev{$\mathrm{Re}=1.73\times10^5$} and $\mathrm{Re}=2.20\times10^5$, where the latter matches the value used in \cite{Slaughter2023}.

\begin{figure}[htbp]
    \centering
    \includegraphics[width=0.7\textwidth]{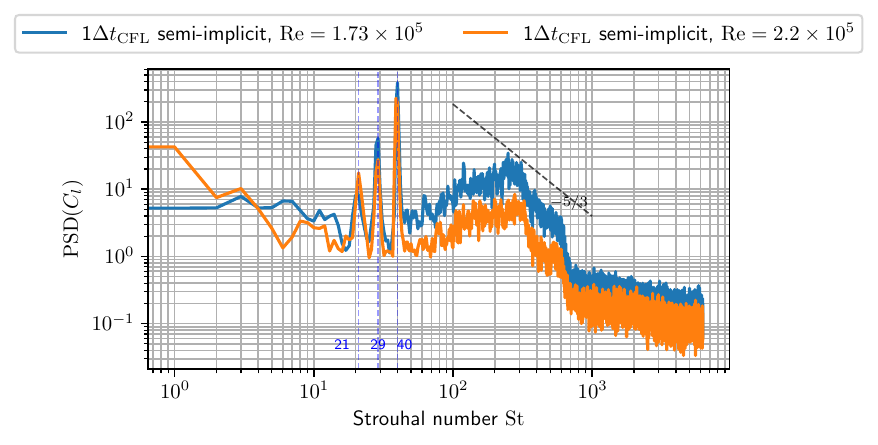}
    \caption{Power spectral density of the total lift coefficient for the semi-implicit scheme comparing the influence of Reynolds number. \rev{The curves are labelled by Reynolds number; $\mathrm{Re}=2.2\times10^5$ is the Reynolds number of \cite{Slaughter2023}.}}
    \label{fig.reynolds-psd}
\end{figure}

Figure~\ref{fig.reynolds-psd} compares the power spectral density of the total lift coefficient for the two Reynolds numbers.
The dominant narrow-band peaks remain located at $\mathrm{St} \in \{21, 29, 40\} $ in both cases, and the high-frequency decay remains qualitatively similar.
This indicates that, for the present setup, the dominant frequency content is only weakly affected by the Reynolds-number change over this range.

In contrast, the time-averaged global forces show a more noticeable sensitivity to Reynolds number, as summarised in table~\ref{tab.reynolds-forces}.
Increasing the Reynolds number from \rev{$1.73\times10^5$} to $2.20\times10^5$ reduces the drag coefficient by $5.1\%$ and increases the lift coefficient by $1.3\%$, consistent with a thinner boundary layer and stronger suction on the wing profiles.
The corresponding force values remain within the range reported by previous studies.

\begin{table}[htb]
    \centering
    \caption{Time-averaged force coefficients and dominant PSD peaks for the two Reynolds numbers considered.}
    \label{tab.reynolds-forces}
    \begin{tabular}{llll}
        Source & $\overline{C}_l$ & $\overline{C}_d$ & PSD peaks in $\mathrm{St}$ \\
        \toprule
        Current study with \rev{$\mathrm{Re} = 1.73 \times 10^5$} & -8.3486 & 0.1891 & $21, 29, 40$ \\
        Current study with $\mathrm{Re} = 2.20 \times 10^5$ & -8.4592 & 0.1794 & $21, 29, 40$ \\
        \bottomrule
    \end{tabular}
\end{table}

\begin{figure}[htbp]
    \centering
    \includegraphics[width=\textwidth]{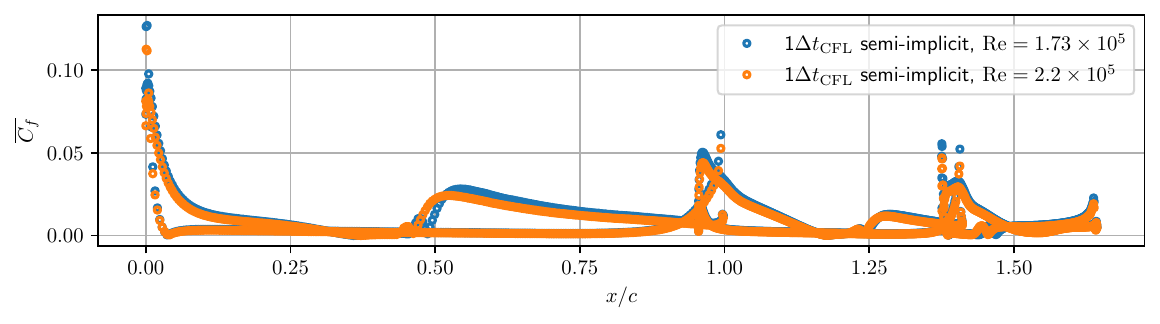}
    \caption{Time-averaged skin-friction coefficients comparing the influence of Reynolds number. Data are averaged in the spanwise direction. \rev{The curves are labelled by Reynolds number; $\mathrm{Re}=2.2\times10^5$ is the Reynolds number of \cite{Slaughter2023}.}}
    \label{fig.reynolds-surface-skinfriction}
\end{figure}

The Reynolds number increase also shifts the transition to turbulence upstream, as shown by the skin-friction coefficient in figure~\ref{fig.reynolds-surface-skinfriction}.
The effect is most visible on the suction side of the main plane and on the pressure side of flap~2.

Overall, the Reynolds number comparison supports the interpretation that the lower Reynolds number used in the main study contributes to differences in mean forces and transition location relative to \cite{Slaughter2023}.
At the same time, it suggests that the dominant spectral peaks observed in the present simulations are not primarily controlled by this Reynolds number variation alone.

\begin{revblock}
\section{Probe locations and transition mechanisms on the flaps}
\label{ap.probes}

Section~\ref{sec.accuracy-scheme-spectra} introduced a turbulent-kinetic-energy
(TKE) probe spectrum as a probe-based analogue of the force-coefficient PSD,
following the same approach already used for the main plane in
section~\ref{sec.verification} (figure~\ref{fig.psd-his-mainplane-mid}).
The main-plane probe (point~\#6 of table~\ref{tab.probe_locations}) sits on
the suction side, \rev{within the chordwise extent of the laminar separation bubble (LSB) reported
in section~\ref{sec.accuracy-scheme-surface} ($x/c \approx 0.43$)}; the flap~1 probe used in
figures~\ref{fig.psd-probe-scheme} and~\ref{fig.psd-probe-dt} (point~\#93 of
the same table) \rev{sits just upstream of flap~1's own LSB ($x/c \approx 1.14$)}. This appendix
extends the diagnostic with a wider spread of points along flap~1's suction
side and with both sides of flap~2, so that the spectral evolution across
the whole pre-transition-to-breakdown region is visible, and a probe is
available on every side that hosts a transition mechanism discussed
elsewhere in the paper -- including flap~2's pressure-side LSB and
suction-side bypass transition, neither of which is otherwise probed.

Table~\ref{tab.probe_locations} lists every probe used in the paper, by wing
element, side and point index: the main-plane suction-side probes, the
flap~1 suction-side probe used for the scheme comparison above, and five
further, evenly spaced points on each of flap~1's suction side and flap~2's
suction and pressure sides, ordered from upstream (leading edge) to
downstream; figure~\ref{fig.probe-geometry} in section~\ref{sec.verification}
shows these locations relative to the wing geometry.
Consistent with a front wing generating downforce, the suction side of every
element is the underside (smaller $z$) and the pressure side the topside
(larger $z$); this is also consistent with the main-plane and flap~2 sides
already identified in section~\ref{sec.accuracy-scheme-surface}
and \ref{ap.reynolds} from the mean surface data.

\begin{table} [hbt!]
   \caption{Probe locations used in figure~\ref{fig.psd-his-mainplane-mid},
   figure~\ref{fig.probe-geometry}, or otherwise referenced in the text, by
   wing element, side and point index (as ordered in the corresponding
   history-point output file).}
   \label{tab.probe_locations}
   \centering
   \begin{tabular}{lllccc}
   \toprule
   Element & Side & Point & $x$ & $y$ & $z$  \\
   \midrule
   Main plane & Suction  & $\# 0$   & -0.8202374 & -0.025 & 0.0700000  \\
   Main plane & Suction  & $\# 2$   & -0.7600429 & -0.025 & 0.0655000  \\
   Main plane & Suction  & $\# 4$   & -0.7506664 & -0.025 & 0.0663000  \\
   Main plane & Suction  & $\# 6$   & -0.7400243 & -0.025 & 0.0675000  \\
   Main plane & Suction  & $\# 8$   & -0.7178520 & -0.025 & 0.0708000  \\
   Flap~1     & Suction  & $\# 107$ & -0.5967520 & -0.025 & 0.1132538  \\
   Flap~1     & Suction  & $\# 93$  & -0.5628000 & -0.025 & 0.1206366  \\
   Flap~1     & Suction  & $\# 86$  & -0.5467067 & -0.025 & 0.1273560  \\
   Flap~1     & Suction  & $\# 76$  & -0.5244402 & -0.025 & 0.1385270  \\
   Flap~1     & Suction  & $\# 65$  & -0.5004225 & -0.025 & 0.1517183  \\
   Flap~2     & Suction  & $\# 126$ & -0.4883012 & -0.025 & 0.1627943  \\
   Flap~2     & Suction  & $\# 117$ & -0.4757957 & -0.025 & 0.1682496  \\
   Flap~2     & Suction  & $\# 108$ & -0.4639274 & -0.025 & 0.1749873  \\
   Flap~2     & Suction  & $\# 98$  & -0.4510663 & -0.025 & 0.1830160  \\
   Flap~2     & Suction  & $\# 89$  & -0.4395649 & -0.025 & 0.1903548  \\
   Flap~2     & Pressure & $\# 11$  & -0.5040965 & -0.025 & 0.1643305  \\
   Flap~2     & Pressure & $\# 23$  & -0.4870277 & -0.025 & 0.1704833  \\
   Flap~2     & Pressure & $\# 34$  & -0.4713039 & -0.025 & 0.1759925  \\
   Flap~2     & Pressure & $\# 46$  & -0.4548900 & -0.025 & 0.1837861  \\
   Flap~2     & Pressure & $\# 58$  & -0.4392995 & -0.025 & 0.1931379  \\
   \bottomrule
   \end{tabular}
\end{table}

Figure~\ref{fig.psd-flap1-suction} shows the TKE spectrum at five points on
flap~1's suction side, using the same probe spacing, Welch parameters and
normalisation as figure~\ref{fig.psd-his-mainplane-mid}; this includes point
\#93, used for the scheme comparison in figures~\ref{fig.psd-probe-scheme}
and~\ref{fig.psd-probe-dt} above.
As on the main plane, a narrow-band peak at $\mathrm{St}\approx40$ emerges and grows
in magnitude moving downstream, and the high-frequency content spreads over
an increasingly broad band -- the same signature of a growing travelling-wave
instability ahead of transition reported for the main plane
(section~\ref{sec.verification}) and attributed by \cite{Slaughter2023} to
the laminar separation bubble.
This confirms, directly in the spectral domain, that flap~1 undergoes the
same LSB-driven transition mechanism as the main plane.
The most upstream point, \#107, sits close enough to flap~1's leading edge to
also pick up a much stronger, broadband imprint of the main-plane trailing
edge shedding through the main-plane/flap~1 gap; \#93 and the other points
further downstream are clear of this direct upstream influence and better
represent flap~1's own boundary-layer development, which is why \#93 rather
than \#107 was chosen for the scheme comparison above.

Figure~\ref{fig.psd-flap2-suction} shows the corresponding spectra on the
suction side of flap~2, where no LSB was reported and the transition was
instead attributed to a bypass mechanism
(section~\ref{sec.accuracy-scheme-surface}, in agreement with
\cite{Slaughter2023}).
Narrow-band peaks at $\mathrm{St}\approx28$ and $\mathrm{St}\approx55$ again grow with
downstream distance, but -- unlike flap~1 -- without the abrupt change in
high-frequency content that marks a separation-driven breakdown: the
spectra broaden smoothly from one probe to the next.
This is consistent with bypass transition being a continuous, distributed
amplification process rather than one tied to a localised bubble.

\begin{figure}[htbp]
    \centering
    \begin{subfigure}[b]{\textwidth}
        \centering
        \includegraphics[width=0.62\textwidth]{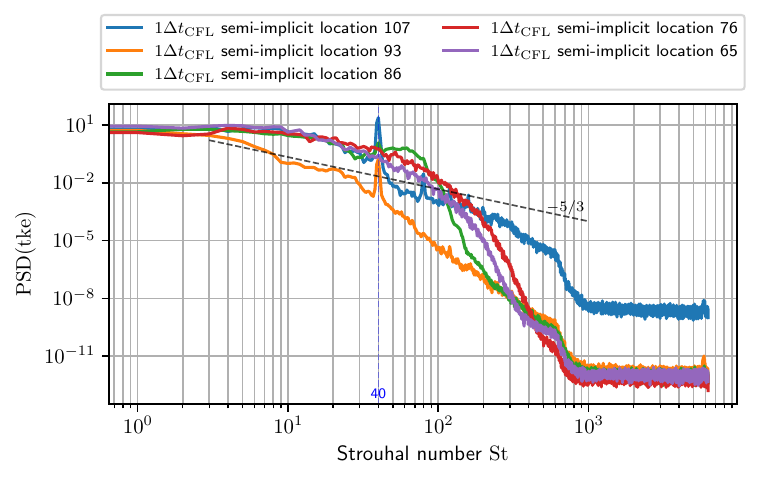}
        \caption{Suction side, flap~1.}
        \label{fig.psd-flap1-suction}
    \end{subfigure}
    \newline%
    \begin{subfigure}[b]{\textwidth}
        \centering
        \includegraphics[width=0.62\textwidth]{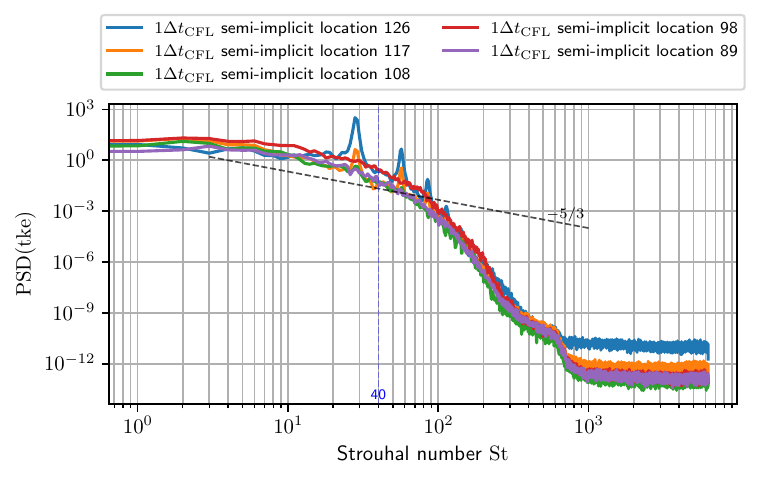}
        \caption{Suction side, flap~2.}
        \label{fig.psd-flap2-suction}
    \end{subfigure}
    \newline%
    \begin{subfigure}[b]{\textwidth}
        \centering
        \includegraphics[width=0.62\textwidth]{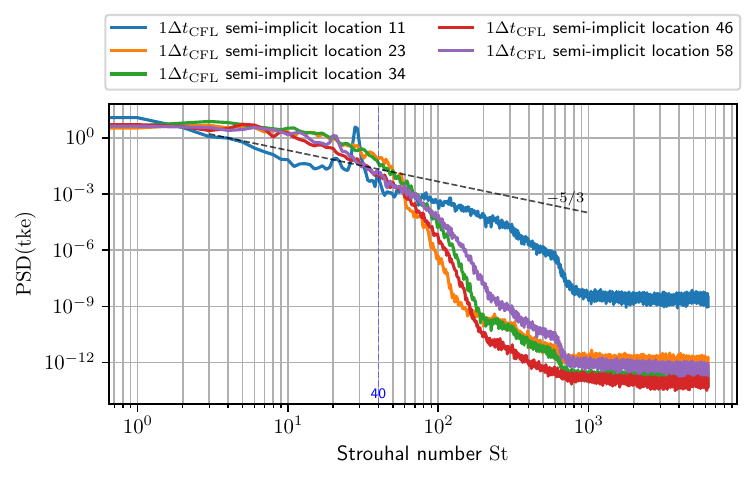}
        \caption{Pressure side, flap~2.}
        \label{fig.psd-flap2-pressure}
    \end{subfigure}
    \caption{Power spectral density of the resolved turbulent kinetic energy
    at five points per side (semi-implicit scheme, $\dtcfl$), ordered
    upstream to downstream as in figure~\ref{fig.probe-geometry}.}
    \label{fig.psd-flaps}
\end{figure}

Figure~\ref{fig.psd-flap2-pressure} shows the pressure side of flap~2, where
a laminar separation bubble was identified from the mean skin-friction
coefficient (section~\ref{sec.accuracy-scheme-surface} and
\ref{ap.reynolds}).
The spectrum at the most upstream probe (\#11, just downstream of the
leading edge) retains broadband energy out to $\mathrm{St}\approx700$, an order of
magnitude further than any other probe considered in this appendix; the
four downstream probes, which lie at and beyond the bubble, collapse onto a
markedly steeper roll-off past $\mathrm{St}\approx100$.
This abrupt change between the first and second probe locates the onset of
the bubble directly in the probe spectra, complementing the mean-flow
evidence from the skin-friction coefficient, and further supports the
picture -- consistent across the main plane and flap~1 above -- that a
laminar separation bubble is associated with a sharp, localised loss of
high-frequency content rather than the smooth broadening seen for bypass
transition.

Taken together, figures~\ref{fig.psd-flap1-suction}--\ref{fig.psd-flap2-pressure}
confirm, in the resolved instantaneous field, the transition mechanisms
inferred from the mean surface data in section~\ref{sec.accuracy-scheme-surface}
and reported for this configuration in \cite{Slaughter2023}: a laminar
separation bubble on the main plane and on flap~1 (both suction side) and on
flap~2 (pressure side), and bypass transition, without a bubble, on the
suction side of flap~2.
\end{revblock}

\bibliographystyle{elsarticle-num}
\bibliography{citations}

\end{document}